\documentclass[%
 aip, cha,
 amsmath,amssymb,
 reprint,%
author-year,%
]{revtex4-1}

\usepackage{graphicx,subfigure}
\usepackage{dcolumn}
\usepackage{bm}
\usepackage{mathrsfs}

\usepackage{breakurl}

\usepackage[utf8]{inputenc}
\usepackage[T1]{fontenc}
\usepackage{mathptmx}
\usepackage{etoolbox}

\usepackage[shortlabels]{enumitem}

\newcommand{\W}{\mathbf W}
\newcommand{\f}{\mathbf f}
\newcommand{\x}{\mathbf x}
\newcommand{\X}{\mathbf X}
\newcommand{\0}{\mathbf 0}

\newcommand{\A}{\mathbf A}
\newcommand{\B}{\mathbf B}
\newcommand{\C}{\mathbf C}

\graphicspath{{figures/}} 

\makeatletter
\def\@email#1#2{%
 \endgroup
 \patchcmd{\titleblock@produce}
  {\frontmatter@RRAPformat}
  {\frontmatter@RRAPformat{\produce@RRAP{*#1\href{mailto:#2}{#2}}}\frontmatter@RRAPformat}
  {}{}
}%
\makeatother

\begin{document}

\preprint{AIP/123-QED}

\title[Intraseasonal atmospheric variability under climate trends]{Intraseasonal atmospheric variability under climate trends}
\author{B. Maraldi}
 \email{bernardo.maraldi@bsc.es}
 \affiliation{%
Department of Physics, Institute for Marine and Atmospheric research Utrecht,
Utrecht University, Utrecht, the Netherlands
}%
\affiliation{Geosciences Department and Laboratoire de M\'et\'eorologie Dynamique (CNRS \& IPSL), Ecole Normale Sup\'erieure and PSL University; Paris, 75005, France}
\affiliation{Barcelona Supercomputing Center, Earth Sciences Department, 08034 Barcelona, Spain}

\author{H. A. Dijkstra}
 \email{H.A.Dijkstra@uu.nl}
\affiliation{%
Department of Physics, Institute for Marine and Atmospheric research Utrecht,
Utrecht University, Utrecht, the Netherlands
}%

\author{M. Ghil}%
 \email{ghil@lmd.ipsl.fr}
\affiliation{Geosciences Department and Laboratoire de M\'et\'eorologie Dynamique (CNRS \& IPSL), Ecole Normale Sup\'erieure and PSL University; Paris, 75005, France}
\affiliation{Department of Atmospheric and Oceanic Sciences, University of California at Los Angeles, Los Angeles, 90095, USA}
\affiliation{Department of Mathematics, Imperial College London; London, SW7 2AZ, United Kingdom}

\date{\today}

\begin{abstract}
 Low-order climate models can play an important role in understanding low-frequency variability in the atmospheric circulation and how forcing consistent with anthropogenic climate change may affect this variability. Here, we study a conceptual model of the mid-latitudes' atmospheric circulation from the perspective of nonautonomous dynamical systems. First, a  bifurcation analysis is carried out under time-independent forcing in order to identify different types of behavior in the autonomous model's parameter space. Next, we focus on the study of the nonautonomous system in which the cross-latitudinal heat flux varies seasonally, according to insolation changes. The forward attractor of the seasonally forced model is compared with the attractor of the autonomous one. The seasonal forcing results in a clear change of the attractor's shape. The summer attractor loses its periodicity, and hence predictability, when the forcing is seasonal, while the winter attractor favors energy transport through one of the model's two wave components. Climate change forcing produces several remarkable effects. Thus, the analysis of the model's snapshot attractor under climate trends suggests that the jet speed does not always follow the sign of the change in equator-to-pole thermal contrast, while the change in the energy transported by the eddies does. Chaotic behavior can be completely suppressed in favor of a regular periodic one and vice-versa. Circulation patterns can change, suddenly disappear, and rebuild. The model's snapshot attractor proves to be a robust tool to study its changes in internal variability due to climate trends, both positive and negative.
\end{abstract}

\maketitle

\begin{quotation}
Edward N. Lorenz introduced a conceptual model of the mid-latitude atmospheric circulation \citep{lorenz84} to investigate the temporal irregularity of the atmosphere and to study the effects of spatially symmetric and asymmetric forcing upon it. 
This model is far from being a detailed and exact representation of the atmosphere, but it has helped test existing theories about the general behavior of the atmospheric circulation in time and space; it has also provided an impetus to study new aspects of this behavior, as done in the present paper.
\end{quotation}

\section{Introduction and Motivation}
\label{sec:intro}


\citet{lorenz84} analyzed the autonomous case of a low-order, mid-latitude  atmospheric model and characterized its behavior by imposing a perpetual season, either summer or winter, through the forcing acting on the system: certain values of the parameters led to coexisting, intransitive periodic solutions, while others induced a chaotic behavior of the system, unveiling the existence of a strange attractor. \\
Given its conceptual simplification of the extratropical atmospheric circulation, this model (L84 from now on) plays a metaphoric role. Nevertheless, dynamical system tools, when applied to such low-order models, prove to be helpful both in characterizing the model's behavior from a merely mathematical perspective and in advancing general insights and an understanding of the processes explaining this behavior. \\ 
Since the publication of the original paper, the L84 model has been studied extensively in both the climate  literature and the mathematical one. \citet{lorenz90} introduced a nonautonomous, seasonal forcing into the L84 model to study the interactions of chaos and intransitivity in the mid-latitude atmospheric circulation. \\ 
Further extensive work by others focused on the bifurcations, the stability and the predictability of the L84 model \citep{shilnikov, Broer, freire2008multistability}. Researching these mathematical properties to characterize the L84 model's behavior proved to be surprisingly interesting for its applications to the study of the atmospheric circulation. \cite{vanveen2001} used the L84 model coupled to a low-order box model for the ocean to study the multiple feedbacks between the atmosphere and the ocean, with the atmosphere acting on a fast time scale and the ocean on a slow one.\\ 
\citet{mangiarotti} compared the L84 model's attractor to that of a model of cereal crops cycles in semiarid regions. 
These authors found that the toroidal structure of the L84 attractor shares certain features with the attractor of their global model for a normalized differential vegetation index. Beside highlighting geometric similarities, the study of the L84 model could thus help one understand the presence of rich dynamics in other areas of applications as well.\\
The reasons to study the L84 model are thus multiple and diverse, each of them focusing on a different perspective on the problem. The system's dynamical properties are quite well understood by now, as far as the autonomous case is concerned. In spite of considerable work over the years and increasing attention to the nonautonomous case, though, there are still many open problems.\\ 
Section \ref{sec:L84} below describes the L84 model from a dynamical systems perspective, touching upon the autonomous case and focusing on the nonautonomous one. We outline why the L84 model is well suited to study low-frequency climate variability, such as subseasonal-to-seasonal (S2S) variability, also referred to as intraseasonal variability \citep{ghil_S2S}. \\
\cite{lorenz84} and \cite{lorenz90} have already provided fine arguments for this usefulness in more process-oriented climate studies.  Here, the spotlight is on the effects of seasonal forcing on the S2S variability by using classical dynamical systems tools, such as bifurcation analysis in Section \ref{sec:system_behaviour}, as well as more recent concepts and methods from the theory of nonautonomous dynamical systems. Most important and helpful among the latter being the {\em pullback attractor (PBA)} and the {\em snapshot attractor} used in Section~\ref{sec:PBA}. \\
Initial applications of the PBA concept to the climate sciences are due to the work of Michael Ghil and collaborators \citep{ghil_PBA, chekroun}. It turns out that one can also define a generalization of forward attraction under certain limiting assumptions on the properties of the time-dependent forcing \citep{Car.Han.2016, Kloeden.Yang.2020}. This restricted form of forward attractors has been called {\em snapshot attractor} in the physical literature \citep{Namenson.ea.1996} and applications thereof to the climate sciences have been initiated  in the work of Tamás Tél and collaborators \citep{Bodai.Tel.2012, Tel.ea.2020}; see Appendix~A. \\
In addition to the seasonal forcing arising from the insolation changes introduced by \cite{lorenz90}, the main purpose of this work is to subject the L84 model to different climate trends that may resemble the effects of global warming and to study the effects of this forcing on the mid-latitude atmospheric circulation and wave patterns. As simple as it is, the system represents an idealized version of this circulation whose main driver, namely the equator-to-pole temperature gradient, will very likely be affected by climate change. \\
The westerly winds between 30$^{\circ}$ and 60$^{\circ}$, both north and south of the equator, are expected to be altered by a trend in the cross-latitudinal temperature gradient. Different effects, though, are observed and predicted at different altitudes in the atmosphere \citep{jetstream}: the Arctic amplification would lead to a reduction of the temperature gradient, as the polar temperatures would warm more than the equatorial ones near the surface. This effect, though, may be quite shallow, concerning only the first few kilometers of the troposphere, and affect only the near-surface meridional temperature gradient. On the other hand, a higher warming due to latent heat release is expected around the tropopause in the tropics, leading to the opposite behavior for the meridional temperature gradient \citep{jetstream}. The latter phenomenon would specifically affect the extratropical jets in the upper troposphere, as opposed to the surface westerlies. \\
Keeping in mind these considerations and the conceptual nature of the model, both alternatives ---  increasing and decreasing cross-latitudinal temperature contrast --- will be considered. Three  different scenarios altogether will be studied: seasonal forcing only, as well as this natural forcing combined with either of two different plausible climate trends. We compute the snapshot attractors for each of the three scenarios. \\
The concept of pullback attraction relies on the fact that measurements happen at the present time, when the forcing has a specific value. When the forcing has been changing, though, over time, past values have to be taken into account; see Appendix~A. It is interesting then to compare the present-time snapshot of the nonautonomous system with the one corresponding to the usual forward attractor for the autonomous case, in which the forcing was held fixed at the same value for the entire evolution of the system. This comparison will clarify the effects of time-dependent forcing  on the climate system in general and on its S2S variability in particular. \\ 
Lastly, a summary of the main results appears in Section \ref{sec:concl}, along with final remarks and comments on future work that takes the issues and methods presented here one step further.

\section{The L84 model and seasonal effects}
\label{sec:L84}
\cite{lorenz84} proposed a system of three ordinary differential equations that conceptually represents the mid-latitude atmospheric circulation.
While in Lorenz's work the model is derived from physical considerations on the  mid-latitude atmosphere's climatological and synoptic dynamics, it is possible to obtain the same system from the analytical approximation of a two-layer quasi-geostrophic flow model \citep{vanveen2003}.
In this section, we summarize both the model derivation and the main processes captured by the L84 model. Furthermore, we define both the model's autonomous and nonautonomous versions, in order to support its use in the study of atmospheric variability. 

\subsection{Model derivation} \label{ssec:derive}
The analytical derivation of the L84 model \citep{vanveen2003} starts with an approximation of the equations for a quasi-geostrophic flow applied to a 2-layer model written in terms of its barotropic streamfunction $\Psi$ and its baroclinic streamfunction $\Theta$ in a cartesian geometry with the $(x,y)$-axes pointing East and North, respectively; see, for instance \cite{Ghil.Chil.1987} or \cite{Kalnay.2003}. \cite{Ghil.Lucar.2020} place this type of fairly simple, idealized model into the more general perspective of a hierarchy of models in the climate sciences \citep{Ghil.2001, Held.2005}.

The main step of the L84 model derivation consists of a Galerkin projection of the equations onto Fourier modes.

\begin{subequations} \label{eq:QG}
	\begin{align}
\Psi(x, y, t) & =  \sum_{m,n} \psi(m, n, t) \exp[i(mkx + nly)], \label{eq:stream} \\
\Theta(x, y, t) & = \sum_{m,n} \theta(m, n, t) \exp[i(mkx + nly)], \label{eq:temp}
\end{align}
\end{subequations}
where $(m,n)$ stand for the zonal and meridional wavenumber, respectively.
After the projection,  \cite{vanveen2003} obtained a six-equation model by only using wave numbers (0,1) and (1,1). The three-equation L84 model is then obtained by keeping only the equations governing the baroclinic streamfunction $\Theta(x, y, t)$. The three-equation model reads as follows:
\begin{subequations} \label{eq:L84}
	\begin{align}
& \frac{d X}{d t} = -Y^2 -Z^2 - aX + aF, \label{l84a} \\
& \frac{d Y}{d t} = XY - bXZ - Y + G,  \label{l84b} \\
& \frac{d Z}{d t} = bXY + XZ - Z.  \label{l84c}
\end{align}
\end{subequations}
Here, $X$ represents the intensity of the zonally symmetric globe-encircling westerly winds, while $Y$ and $Z$ represent the cosine and sine phases of a chain of superposed large-scale eddies, or waves. These waves are crucial in the poleward heat transport \citep{lorenz1967} and, in the model, this transport occurs at a rate proportional to the square of their amplitudes, $Y^2 + Z^2$. 

\subsection{Autonomous case} \label{ssec:auto}

All the variables are scaled so that the unit of time $t$ is 5 days, roughly the time scale for the eddies to damp. Therefore, by tuning the parameter $a$ in Eq.~\eqref{l84a}, it is possible to determine whether the westerlies damp more or less rapidly than the eddies. The parameter $b$ in Eqs.~(\ref{l84b}, \ref{l84c}) defines the time scale of displacement of the eddies due to the current, while $F$ and $G$ are external forcing terms, namely the cross-latitude external heating contrast and the asymmetric forcing arising from land-ocean heating contrast, respectively. Note that, for $G = 0,$ the model is perfectly symmetric with respect to an interchange between $Y$ and $Z.$\\ 
The model is clearly of the forced-dissipative type, like the \cite{Lorenz.1963} convection model, and has a globally attracting set \citep{Ghil.Chil.1987}; see Appendix~B for details. The linear terms are dissipative and the quadratic ones conservative. The dissipation balances the forcing and the irregularity arises from the fight between the forcing and the dissipation within a finite volume in phase space, in the presence of instability.\\
In the model, it is the thermal forcing $F$ that acts to change the equator-to-pole temperature gradient $X$. When $F$ does not depend on time, the system is said to be {\em autonomous}. In the autonomous case, the forcing value $F=6$ models a perpetual summer season, with a weaker jet intensity $X,$ while $F=8$ corresponds to a perpetual winter, with a stronger jet \citep{lorenz84}. 

\subsection{Nonautonomous case} \label{ssec:nonauto}
\begin{figure}[t!]
    \centering
    \includegraphics[width=0.5\textwidth]{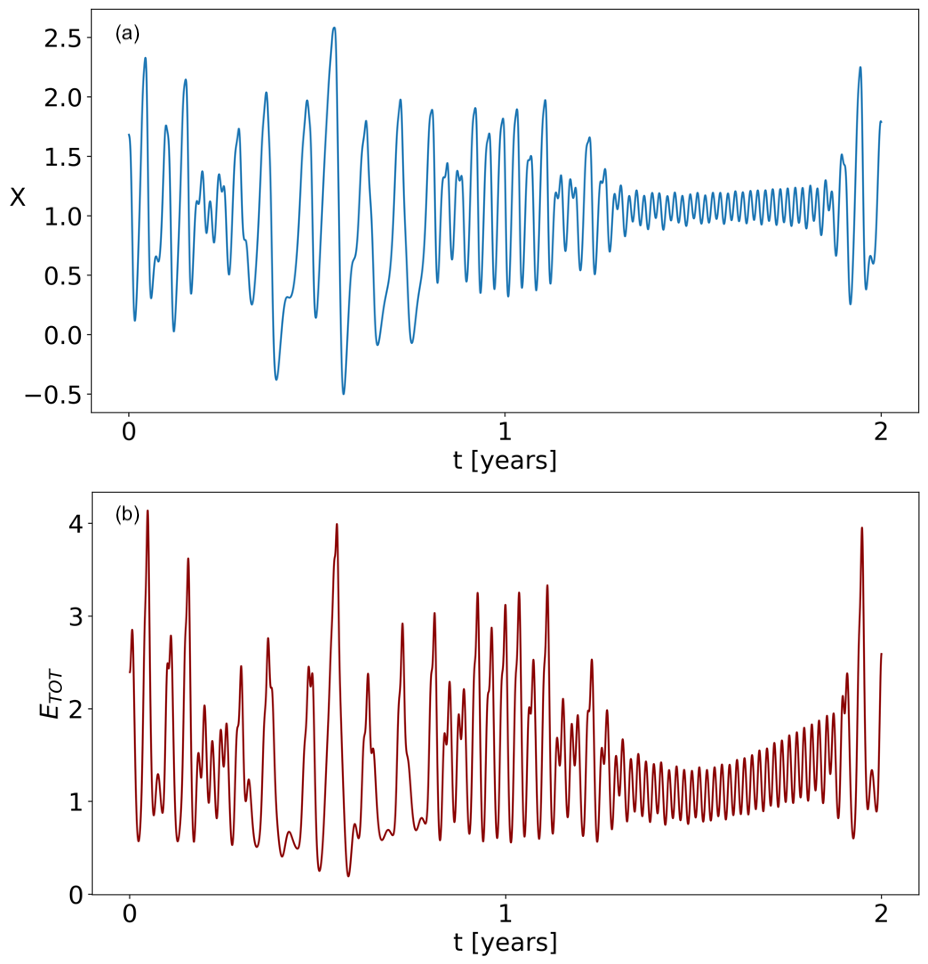}
    \caption{Final 2 years of a 20-year simulation with a seasonal forcing of $F(t) = F_0 + A \cos(\omega t), \ \omega = {2\pi}/{\tau}$. The parameter values are: $F_0 = 0, A = 2, a = 0.25, b = 4,$ and $G = 1.$ These values were chosen in order to achieve different model behavior between the seasons: summers are roughly periodic and winters are chaotic. (a) Jet intensity $X = X(t)$ over time. Two different types of behavior arise for the summer season, \textit{active} for the first year and \textit{inactive} for the second year. (b) Energy $E_{TOT}$ of the system over time. The active summer displays energy fluctuations with higher amplitudes and lower frequency, while the opposite happens for the inactive summer). }
    \label{fig:timeseries}
    \vspace{-10pt}
\end{figure}

In fact, though, the meridional temperature gradient varies smoothly according to the change of the seasons, from being higher in winter to being lower in summer.  
Therefore, one has to consider its explicit dependence on the time to study seasonal and climatological effects. When $F$ explicitly depends on the time $t$, the system is said to be {\em nonautonomous}. In the latter case, the forcing term appears as:
\begin{equation}\label{seasonal}
    F(t)=F_0 + A \cos(\omega t), \; \omega = {2\pi}/{\vartheta}.
\end{equation}
Here $A$ is the amplitude of the seasonal oscillation, while $\vartheta$ equals one year, so that $\vartheta \ = 73$ since the model's time unit is five days, and the year is taken to have exactly 365 days.\\ 
Note that no thermal inertia is taken into account in this model and the solar heating is the direct source of energy, with the maxima and the minima of the forcing assumed to occur at the start of the year and in the middle. 
In reality, it is the underlying ocean and land that are the main source of heat, and a lag exists between the solstices and the maxima of heat provided by the land and the oceans. \\
To take a global climate trend into account, we add a linear term to the constant term $F_0$. This additional term added to Eq.~\eqref{seasonal} will be linearly decreasing when focusing on the near-surface flow, where a moderate westerly wind exists. To the contrary, we add a linear increase to $F_0$ when considering the higher altitudes at which the jet streams develop, with their higher wind velocities. \\
Specifically, for climatic trends, $F_0$ in Eq.~\eqref{seasonal} will be replaced by $F_1 = F_1(t)$, as follows:
\begin{equation} \label{F(t)}
    F_1(t)=
    \left\{
\begin{aligned}
    & F_0 ,\ t < 10 \vartheta; \\
    & F_0 \pm \alpha \frac{(t-10 \vartheta)}{T},\ t \ge 10 \vartheta. \\
\end{aligned}
\right.
\end{equation}
At $t=0$, the starting value is ${F_1}=7$ and $T=100$~years. The value of the slope, $\alpha = 2/T$, was chosen to be high enough to guarantee that the forcing will assume values consistent with different types of model behavior.
In the expression \eqref{F(t)}, a 10-year time span is left for the system to reach a stationary climate. Although one could argue that a longer pre-industrial time interval should be prescribed, \cite{drotos} have shown that the L84 model's convergence time to its attractor equals roughly 5 years.\\ 
For the sake of completeness, one could consider additional types of forcing that also change seasonally. For instance, $G$, which depicts the heat contrast between land and ocean, could also be time dependent but, throughout this work, we have used $G=1$, to avoid unnecessary complications at this stage.
The other parameters are also kept fixed at their traditional values, $a = 0.25$ and $b = 4$, except in Section~\ref{sec:system_behaviour}, where the dependence of the system on the prescribed but time-independent value of $a$ is investigated. \\
The model's parameter values differ slightly from those based on observations and used in more detailed models. Nevertheless, the main physics described by the model is true to that found in high-end model simulations, and all the model variables are just a scaled version of the variables from the original truncated system: for instance, the forcing $F$ is a scaled, nondimensional version of the observed cross-latitudinal temperature gradient. \\
In addition, a comparison between the bifurcation diagrams of the original quasi-geostrophic model and the L84 model 
confirms that the latter provides a good qualitative representation of the underlying phenomenon it synthesizes, namely, the interaction between westerly jets and the superimposed baroclinic waves. The inescapable limitations imposed by low resolution in such a comparison are discussed, for instance, in \citet[Sec.~5.3]{Ghil.Chil.1987}. \\
Keeping in mind the time scaling of the variables, numerical simulations have been carried out to explore the behavior of the model over time. Specifically, we used a fourth-order Runge-Kutta numerical scheme with a time step $\delta t = 0.025$, which corresponds to 3 hours in dimensional time, to integrate the system. Typical results of a 20-year--long model simulation with seasonal forcing are shown in Fig.~\ref{fig:timeseries}.

The zonal-wind variable $X = X(t)$ is shown in Fig. \ref{fig:timeseries}(a). The figure shows two consecutive years of a 20-year simulation, starting from the winter solstice, with initial conditions ${\X(t = 0)} = (2, 1, 0)$, where ${\X} = (X, Y, Z)^T$ is the coordinate vector in phase space. Recalling the convention of no lag in the seasons with respect to the forcing, and no transition seasons, summers last roughly from $t = n + 1/4$ years to $t = n + 3/4$ years.  \\
As noted by \cite{lorenz90}, the system exhibits a bimodal behavior during the summer seasons, with the flow displaying either slow oscillations with high amplitude or faster oscillations of smaller amplitude. The two types of behavior have been labeled "active" and "inactive" summers. It is the amplitude of the oscillations during a season that distinguishes between the two. \\
It is interesting to relate this bimodality to the behavior of the waves and of the total energy of the system, defined as $E_{TOT} = 0.5(X^2+Y^2+Z^2)$. Figure~\ref{fig:timeseries}(b) shows the energy of the system over the same time interval. Note that, during the active summer, the energy of the system undergoes higher-amplitude oscillations than in the inactive one, as is the case for the zonal winds. \\
The figure shows that the two possible types of behavior take turns over the years. At the same time, it is possible for one of the two types of summer to prevail for many consecutive years (not shown). It is the chaotic winter that randomly resets the initial conditions for the following summer and that leads to interannual variability. \\
This means that, during an active summer, the interaction of the zonal flow with the waves is stronger and waves are constantly pumping and extracting higher quantities of energy from the flow. During the inactive summers, there is little energy in the system and the interaction between the waves and the zonal flow is much faster, leading to smaller oscillations with higher frequency: in fact, the number of oscillations is almost double that during the active summer. Therefore,one can look at the amplitude of the total energy fluctuations as an index of eddy activity. 
\\
Moreover, it is important to highlight that the oscillations in $X(t)$ have a mean period of roughly 20--30 days, although the behavior is chaotic, especially during the winter. \\ 
The frequency of these oscillations suggests that the L84 model, within certain parameter ranges, could be a suitable candidate to investigate the effect that climate change may have on atmospheric S2S variability. This type of variability is generally associated with periodicities that span a range from a few weeks to one or two months \citep{ghil_S2S, vitart2019introduction}. Specifically, for the extratropical, mid-latitude flows, the dominant period of the variability is about 40--50 days, which is somewhat longer than what is found for the L84 model, but still broadly compatible; see also \citet[Sec.~E.3]{Ghil.Lucar.2020}.

\section{Autonomous system behavior and its parameter dependence}
\label{sec:system_behaviour}
We study in this section the dependence of the autonomous model's behavior on the parameters $a$ and $F$. The L84 model's bifurcation tree has been studied before \citep{shilnikov, Broer, vanveen2003}. Here, the steady states of the system are studied along with the transient behavior.

\subsection{Steady states and bistability} \label{ssec:2_fold}
The steady states of the system are found by setting the right-hand side of equations~(\ref{eq:L84}a,b,c) 
equal to zero. The resulting equations are:
\begin{subequations} \label{eq:FPs}
	\begin{align}
& Y = (1-X)G/(1-2X-(1+b^2)X^2), \label{eq:FPa} \\
& Z = bXG/(1-2X-(1+b^2)X^2), \label{eq:FPb} \\
& a(F-X)(1-2X-(1+b^2)X^2)-G^2=0. \label{eq:FPc} 
\end{align}
\end{subequations}
By solving first equation \eqref{eq:FPc} for $X$, it is straightforward to obtain the remaining two coordinates from Eqs.~(\ref{eq:FPa}, \ref{eq:FPb}). 
\begin{figure}
    \includegraphics[width=0.96\columnwidth]{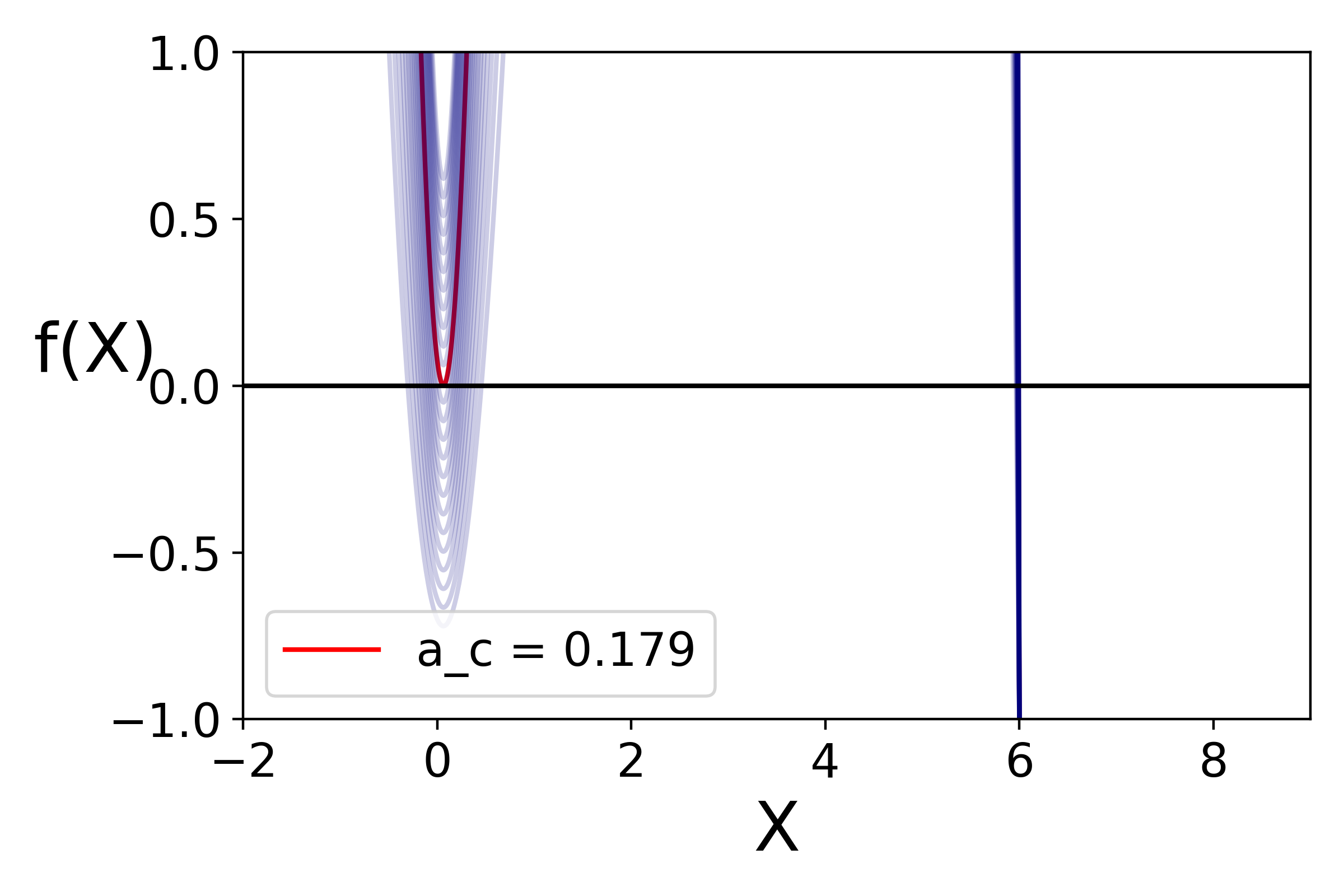}
    \caption{Graph of the cubic polynomial $f(X) = a(F-X)(1-2X-(1+b^2)X^2)-G^2$ for different values of the control parameter $a$, with $F = 6$ and $G = 1$. The graph's intercepts with the abscissa coincide with the $X$-coordinate of the system's steady states. When $a$ crosses the critical value $a_c$, the number of steady states changes from 3 to 1, which indicates the occurrence of a saddle-node bifurcation. }
    \label{fig:polynomial} 
\end{figure}
When $a$ is the control parameter, $F$ is kept fixed at the value $F=6$, which corresponds to perpetual summer, although any other value would yield an analogous result (not shown). 
Recall that $G \equiv 1$ and $b\equiv 4$ throughout this paper. \\
The cubic polynomial $f(X)=a(F-X)(1-2X-(1+b^2)X^2)-G^2$ is plotted in Fig.~\ref{fig:polynomial} for different values of the wave damping coefficient $a$, while $F \equiv 6$. 
The figure shows that the system goes from having 3 to only 1 real steady state as the parameter $a$ crosses a threshold $a_c.$ 
A more detailed analysis shows that $a_c$ marks one of the two critical thresholds of a double-fold — or back-to-back saddle-node — bifurcation, the other one being $a = a_d = 0.00189$. We also find that, for $a = a_H=0.0113$, a Hopf bifurcation occurs. \\
Given the shape of the curve, one can analytically compute the value for which the minimum of $f(X)$ changes sign and, therewith, find the critical value $a_c$. Computing the derivative of $f(X)$ and setting it equal to zero yields:
\begin{equation}
\begin{split}
    f'(X) = &X^2(-3-3b^2) + \\
            &X(2F+2b^2F+4)-2F-1 = 0,
\end{split}
\label{der}
\end{equation}
where $a$ cancels out. The solution for the minimum is $X_c = 0.06$ for $F = 6$ and, from Eq.~\eqref{eq:FPc}, it then immediately follows that
\begin{equation} \label{eq:a_crit}
    a_c = \frac{G^2}{(F-X_c)(1-2X_c +(1+b^2)X_c^2)} = 0.179.
\end{equation}

\subsection{Successive bifurcations, Lyapunov exponents, and basins of attraction}
\label{sec:subsection_bif}
\begin{figure}[h]
	\hspace{-10pt}
    \includegraphics[width=0.5\textwidth]{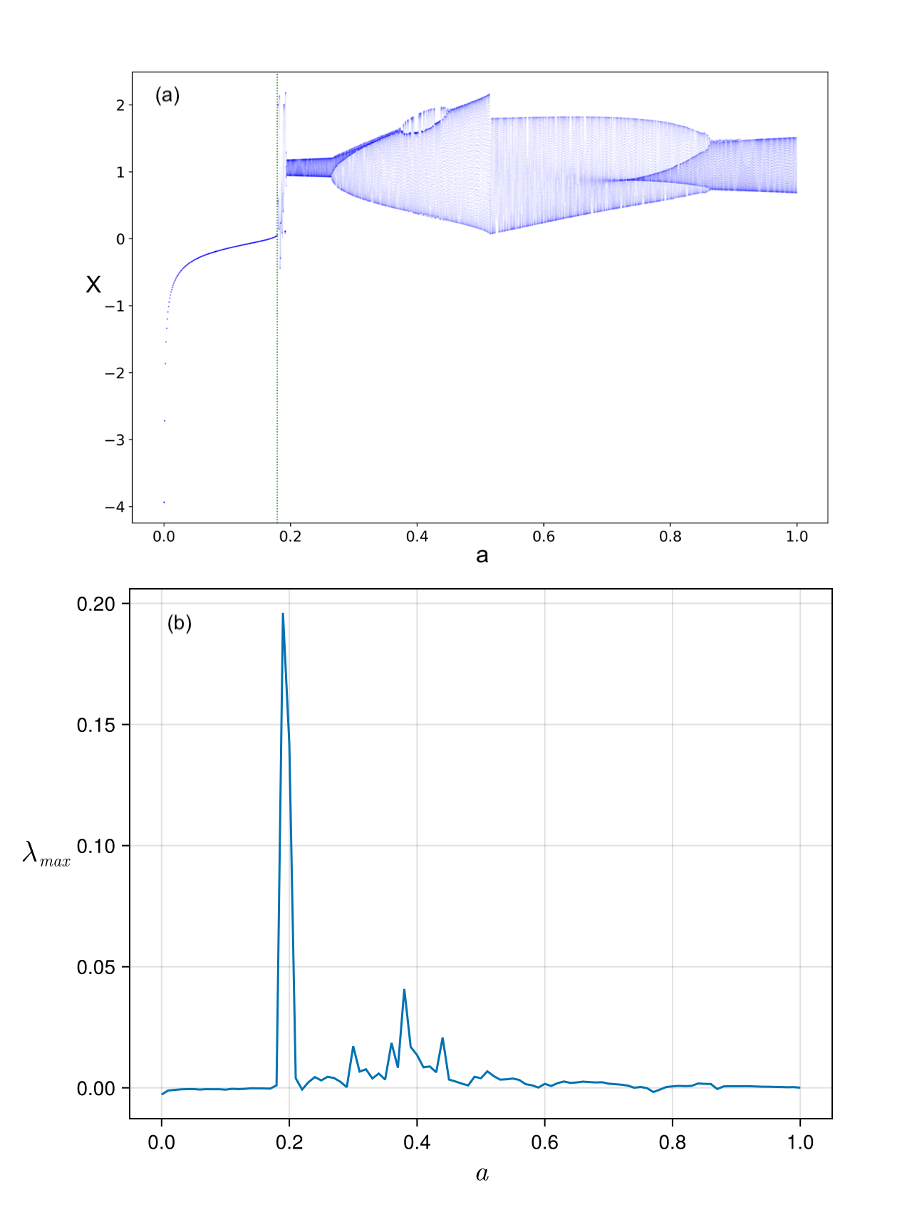}
    \caption{Partial bifurcation diagram for the autonomous case of the L84 model, with $a$ as the control parameter and $F = 6$. (a) Here $a$ assumes 1~000 equidistant values in the interval $[0,1]$. For each value of $a$ the last 100 points of the time series of $X$ are shown.
    A dashed vertical line highlights the $a = a_c$ value at which the saddle-node bifurcation highlighted in Fig.~\ref{fig:polynomial} happens. (b) Largest Lyapunov exponent for each $a$-value.}
    \label{fig:bifurcation_a}
\end{figure}

\begin{figure}[h]
 	\includegraphics[width=0.47\textwidth]{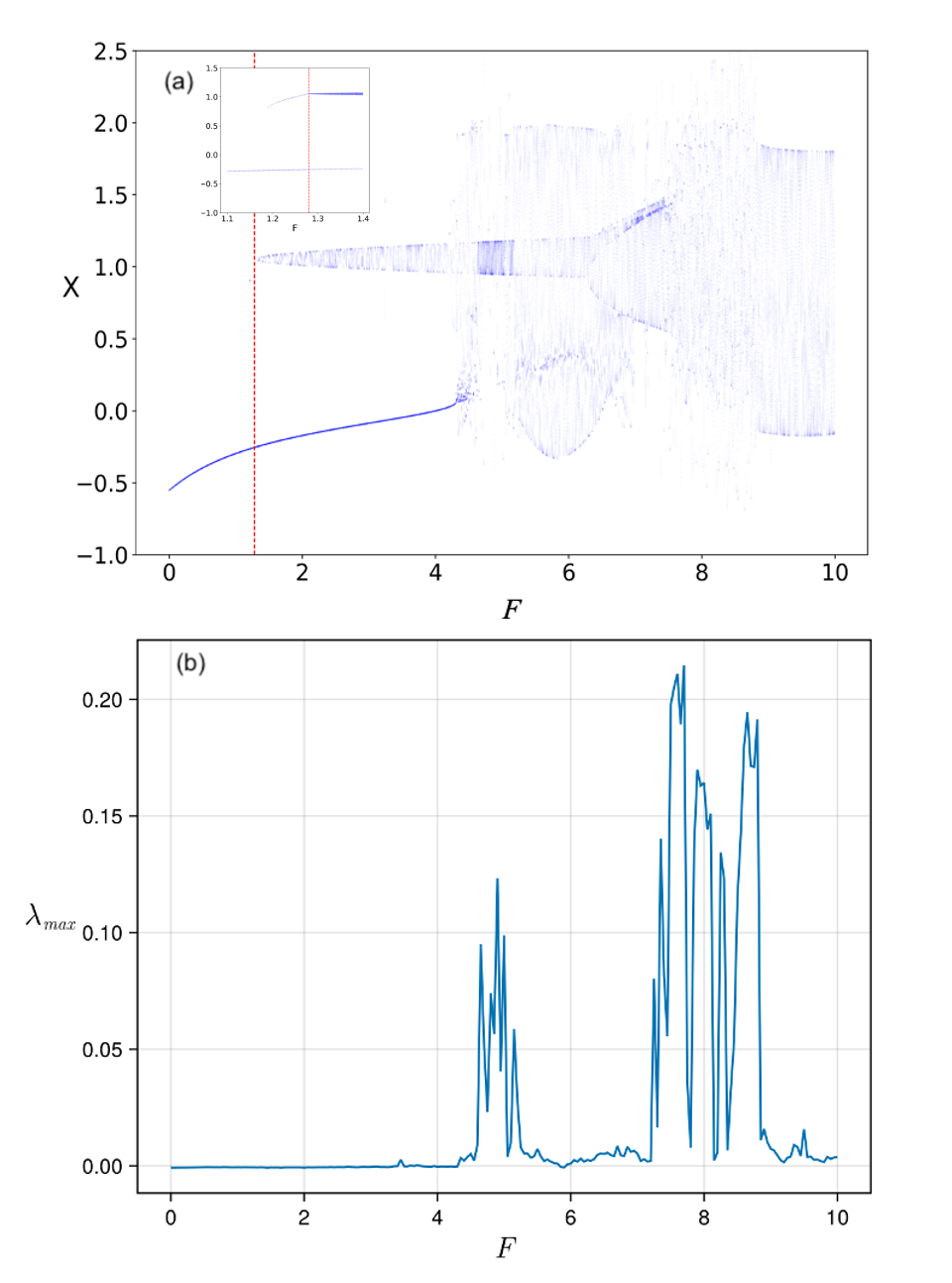}
 	\caption{Partial bifurcation diagram for the autonomous L84 model with $F$ as a control parameter and $a = 0.25$. (a) Here $F$ assumes 1~000 equidistant values in the interval $[0,10].$ 
 		In this case, for each value of $F$, we use 100 runs, each of which starts with a random initial condition in the cube $D \subset \mathbb R^3$, and it is the last 100 points of the 100 runs that are shown in the figure. The inset focuses on the coexistence of a stable stationary solution and a stable limit cycle near the first saddle-node bifurcation and the Hopf bifurcation. 
 			(b) Largest Lyapunov exponent for each $F$-value.}
 	\label{fig:colonna}
 \end{figure}

\begin{figure}[h]
\hspace{-10pt}
    \includegraphics[width=0.47\textwidth]{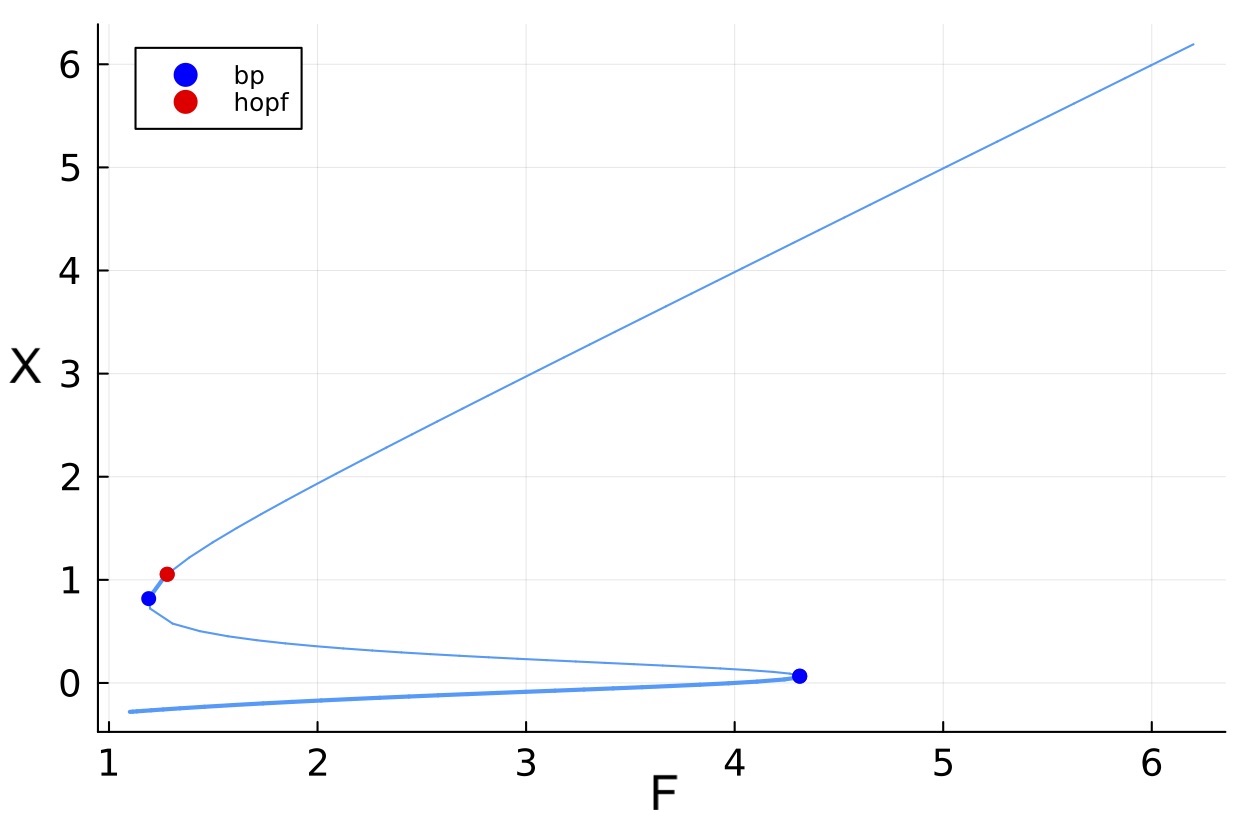}
    \caption{Bifurcation diagram for the autonomous L84 case with $F$ as the control parameter and $a=25$. The two blue points denote a saddle-node bifurcation each, while the red one marks a Hopf bifurcation.}
    \label{fig:bif_kit}
\end{figure}

\begin{figure}[h]
\vspace{-25pt}
\hspace{-10pt}
    \includegraphics[width=0.47\textwidth]{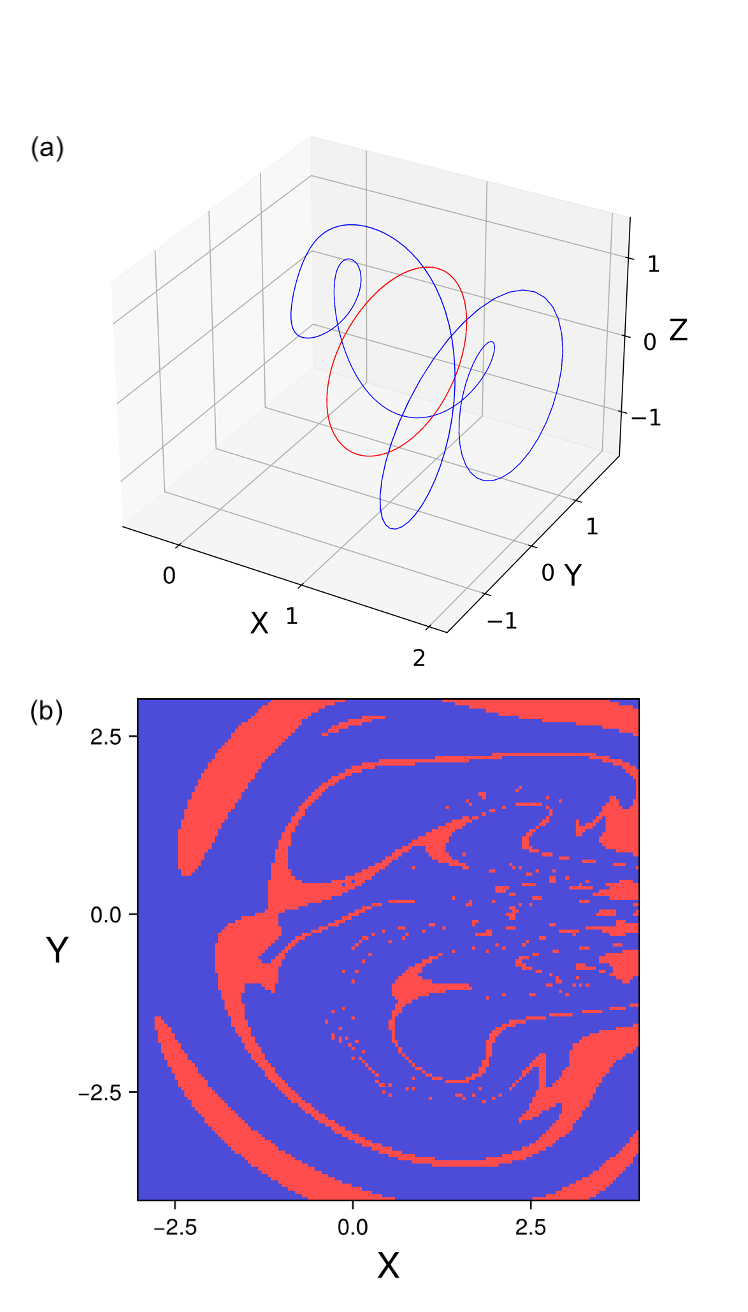}
    \caption{Numerical simulation with $10^{4}$ random initial conditions in the cube $D \subset \mathbb R^3$ used in Fig.~\ref{fig:colonna}, for a perpetual summer with $F \equiv 6$ and with $a = 0.25$. 
    	(a) Axonometric projection of two separate limit cycles, blue and red. (b). $(X, Y)$-plane cross section of the attractor basins of the two limit cycles.}
    \label{fig:basins}
\vspace{-15pt}
\end{figure}

The bifurcation analysis of the autonomous system in this section is carried out with respect to the control parameters $a$ and $F$, separately, by using two different procedures. The first procedure, when keeping the forcing $F$ fixed, is analogous to that of \cite{Broer}: 1~000 values of $\{a_k = k \Delta a: k = 1, \ldots, 1~000\}$, with $\Delta a = 1/1~000,$ have been selected in the parameter interval $[0,1]$.\\
A numerical simulation is then carried out for each value of the damping parameter $a$ and the last 100 points of the trajectory of $X(t)$ are shown in Fig.~\ref{fig:bifurcation_a}. 
In this case, the initial condition for the first value of $a$ was ${\W(t = 0)} = (2, 1, 0)$, and the final point of each simulation was set to be the starting condition for the following one. The figure confirms a change of behavior for the value $a_c = 0.179$ found analytically in \eqref{eq:a_crit}. \\
This procedure allows one to track a branch of steady-state solutions but it is not best suited to capture all possible types of behavior. Therefore, in the case of $F$ being the control parameter, we used an ensemble of initial conditions, in order to capture multiple types of behavior. 
The initial conditions are chosen randomly without replacement in the cube $D \subset \mathbb{R}^3$ defined as $D = \{(X,Y,Z) \vert -3  \le X \le 3, -3 \le Y \le 3, -3 \le Z \le 3 \}.$ 
In this case, we had no analytical results for additional types of bifurcations, so that only the results of the numerical study are reported herewith. \\
These results 
are shown in Fig.~\ref{fig:colonna}a. The figure clearly indicates that different types of solutions exist for the same value of $F$. Specifically, one also finds oscillatory behavior and we investigate the existence of Hopf bifurcations that could generate such limit cycles. To do so, we linearize the system~\eqref{eq:L84} around the steady state of interest at a given $F$-value \citep{Kuznetsov.1995, dijkstra2013nonlinear} and track a 
pair of conjugate eigenvalues as their real part crosses the value zero
 \citep{guckenheimer, Ghil.Chil.1987}. 
 Our bifurcation analysis indicates that such a Hopf bifurcation, at which the steady state loses its stability and transfers it to a limit cycle, occurs when $F$ crosses the critical value $F_H = 1.28$ at $a = 0.25$, in agreement with previous work \citep{shilnikov, vanveen2003}. \\
To validate these results we implemented a second procedure, using the continuation algorithms provided by BifurcationKit.jl \citep{veltz:hal-02902346}, which confirmed the presence of a Hopf bifurcation at $F_H$ along with a double-fold bifurcation 
with critical thresholds at $F_1 = 1.19$ and $F_2 = 4.31$. The results are shown in Fig.~\ref{fig:bif_kit}. \\
The dependence of the autonomous system's largest Lyapunov exponent on the damping parameter $a$ and the forcing parameter $F$ is shown in Figs.~\ref{fig:bifurcation_a}b and \ref{fig:colonna}b, respectively, for the region of the parameter space where the system has been previously studied. Recall that the values $F=6$ and $F=8$ have been associated heretofore with a regular, periodic summer and a chaotic winter, respectively. In general, the system starts to display an oscillatory behavior around $F \simeq 2$, depending on the value of $a.$
Figure~\ref{fig:bifurcation_a}b clearly shows that irregular behavior can only occur in summer for a limited range of $a$ values, with strong instabilities in an even more limited range. \\
Intervals of regular periodic and chaotic behavior alternate as the forcing $F$ changes in Fig.~\ref{fig:colonna}b. Overall, instabilities are much stronger and extend over a larger parameter range than in Fig.~\ref{fig:bifurcation_a}b, essentially starting at $F \simeq 4$ for $a = 0.25$.  \\
 It is also important to highlight the existence of multiple attractors for specific values of $F$. Figure~\ref{fig:basins}a shows the coexistence of two separate closed orbits for $F=6$. To study the system's basins of attraction, we used again an ensemble of initial conditions in the cube $D \subset \mathbb{R}^3$, as we did in Fig.~\ref{fig:colonna}. \\
Figure~\ref{fig:basins}(a) shows two limit cycles obtained for perpetual summer ($F \equiv 6$), one with a purely elliptic structure, a smaller amplitude, and a shorter period (red closed curve) than the other  one (blue closed curve). This result is in agreement with those of \cite{lorenz90}, as well as with those in Fig.~\ref{fig:timeseries} herein, and contributes to a better understanding of both. The periods are 7.3 days for the red limit cycle and 35.1 days for the blue one. For orientation purposes, 7 days is comparable to the life cycle of baroclinic eddies in the atmosphere \citep{Kalnay.2003}, while 35 days is comparable to the periodicity of barotropic mid-latitude intraseasonal oscillations \citep{ghil_S2S}.\\
The basins of attractions of these two limit cycles are shown in Fig.~\ref{fig:basins}(b). In this case, as well as for lower values of $F$, each of the two limit cycles in the figure's panel (a) has an attractor basin with an apparently fractal structure \citep{Grebogi.ea.1987}. 
On the other hand, when the system exhibits a chaotic behavior for perpetual winter $F \equiv 8$, a single global attractor, and consequently a single basin of attraction, is found. The libraries DynamicalSystems.jl and Attractors.jl \citep{Datseris2018, Attractors.jl} were used for the computation of the Lyapunov exponents in Figs.~\ref{fig:bifurcation_a} and \ref{fig:colonna}, as well as of the system's attractor basins in Fig.~\ref{fig:basins}. \\

\section{Pullback and forward attraction in the presence of time-dependent forcing}
\label{sec:PBA}

In the study of nonautonomous dynamical systems, the traditional concept of forward attraction does not work in certain cases, as the limit $t \rightarrow + \infty$ is not well defined, especially when the forcing pushes the system to blow up in this limit \citep{Car.Han.2016, Kloeden.Yang.2020}. A new concept has to be proposed to replace it: the attraction is considered after the integration of a large ensemble of initial conditions that converges to a snapshot attractor \citep{Namenson.ea.1996, drotos} or to a pullback attractor,  hereinafter PBA \citep{Crauel.Flan.1994, ghil_PBA}. 

\subsection{Pullback, forward and uniform attractors} \label{ssec:PBA_v_Fwd}
The key property of a PBA is that, rather than observing the asymptotic state of the system in the remote future, $t\to + \infty$, the observation occurs at the present instant $t$, supposing that the system has evolved from an ensemble of initial conditions set at a remote initial time $t_0 = s \to - \infty$. In numerical practice, the initial state does not need to be asymptotically far in the past \citep{chekroun, Pierini.Ghil, charo}.
More is said about PBAs, snapshot and {\em uniform} attractors in Appendix~A. \\ 
Within the setting of the random version of PBAs, also called {\em random attractors}, \cite{Flandoli.ea.2022} have shown that there is a rigorous way of defining two time scales, a "macroscale" corresponding to climate, and a "microscale" that corresponds to weather. 
In the case of the L84 model, we shall take the microscale to be simply $t$, while the macroscale $\tau$, say, will be that of a month, so that one can still have the climate of the model change within a season, as well as from one year to the next. \\
Note that using the PBA concept requires the use of an ensemble of initial conditions, rather than using a single realization, as done in this paper up to this point. Moreover, \cite{drotos} have shown that the snapshot attractor for the L84 model with purely seasonal forcing becomes periodic after a sufficiently long convergence time of $t_c \simeq 5$~years \citep{drotos}. 
Therefore, the choice of which year to use when studying the periodic attractor for sufficiently large $t > t_c$ is arbitrary.  \\
As explained in Appendix~A, one can use the snapshot attractor approach to study the L84 model's seasonal forcing, as done previously by \cite{drotos}. But it would become necessary to use the pullback approach to study monotonically increasing or decreasing forcing and the combination of the seasonal and monotonic forcing if we were really interested in pursuing the monotonic forcing all the way to very large times. \\
Fortunately, \cite{Anguiano.Car.2014} have discussed the possibility of the L84 attractor's being uniform \citep{Haraux.1991,Vishik.1992}, given its uniform dissipativity for $a$ constant, and for a bounded forcing of a fairly general kind. As we are not pursuing herein the climate system's forward limit all the way to blow up, we may assume that the attractor is uniform, having checked that, numerically, the pullback and forward approach yield the same results. This equality is a necessary but not sufficient condition for the existence of a uniform attractor. To distinguish the nonautonomous forward attractor from the autonomous one, we adopt here the snapshot terminology used in the physical literature \citep{Namenson.ea.1996,Tel.ea.2020}. \\
The numerical tests we conducted were for periodic, as well as for aperiodic but bounded forcing. In these tests, the attractor of the L84 model was computed at fixed $t$ starting from compact sets of initial data at $s < t$ with $|t - s|$ increasing; see Appendix A herein and  \citet[Fig.~4]{charo}, as well as  \citet[Fig.~7]{ghil_PBA}. \\
These tests confirmed the similarity of the attractors obtained using this PBA approach with those obtained using the simpler snapshot approach of \cite{drotos}. In particular, our results confirmed the convergence time of $t_c \simeq 5$~years of the latter authors. Given the less laborious character of the snapshot methodology — which only requires forward integration once convergence of initial ensembles to an attractor has been confirmed — we have preferred to use the latter throughout the rest of the paper. \\
In order to study the snapshot attractor of the L84 model, an ensemble of  
$N_{\rm R}$ random initial conditions was initialized in the cube $D \subset \mathbb R^3$ again, as done in obtaining Fig.~\ref{fig:colonna}. Every simulation was carried out with these same initial conditions and the starting time at $s = 0$, while the parameter values were set as before to be $a=0.25,$ $b=4$ and $G=1$. \\
When considering only the seasonal cycle, the forcing is given by Eq.~\eqref{seasonal}, and $N_{\rm R} = 5 \cdot 10^{4}$. The climate trends are accounted for by the additional terms in Eq.~\eqref{F(t)} and, in this case, we used a smaller ensemble of $N_{\rm R} = 10^{4}$, since much longer runs were needed. \\
In the following subsections, we focus on the comparison between the autonomous version of L84 and the nonautonomous one, in order to understand the effects of external time-dependent forcing. We compare specifically the model response to a fixed, constant forcing and to a forcing that explicitly depends on time. 
In particular, the attractor for the nonautonomous case --- once with seasonal forcing alone and once with a linear trend added --- is compared with the forward attractor of the autonomous case. \\
For the purposes of such a comparison, specific snapshots of the uniform attractor have to 
be chosen. In the case with seasonal forcing alone, two different instants are considered, namely at the observation time $t_1$ at which $F(t_1) = 6$ and at $t_2$ at which $F(t_2) = 8$. Recall 
that in the autonomous case, in which a perpetual forcing is imposed, these values of the forcing correspond to a periodic and a chaotic behavior, respectively. In the nonautonomous case, this cannot necessarily be stated only by visual inspection of the time series, since the two types of behavior seem to interlace and so a more careful study was carried out. \\
Note that, in fact, $F=6$ and $F=8$ are not the extremal values of the forcing in Eq.~ \eqref{seasonal}. It would be equally reasonable to take any other time instant with different values of the forcing in the comparison with periodic forcing. These two $F$ values were chosen because the autonomous model behavior had been explored extensively at these values.
For the case of a climate trend being imposed, several instants within a seasonal cycle were used to obtain a set of snapshots. In this case, the effect of a climate trend is reflected in a change of the attractor's structure, as shown below. \\
\begin{figure*}
  \begin{minipage}{0.45\textwidth}
    \centering
    \includegraphics[width=\textwidth]{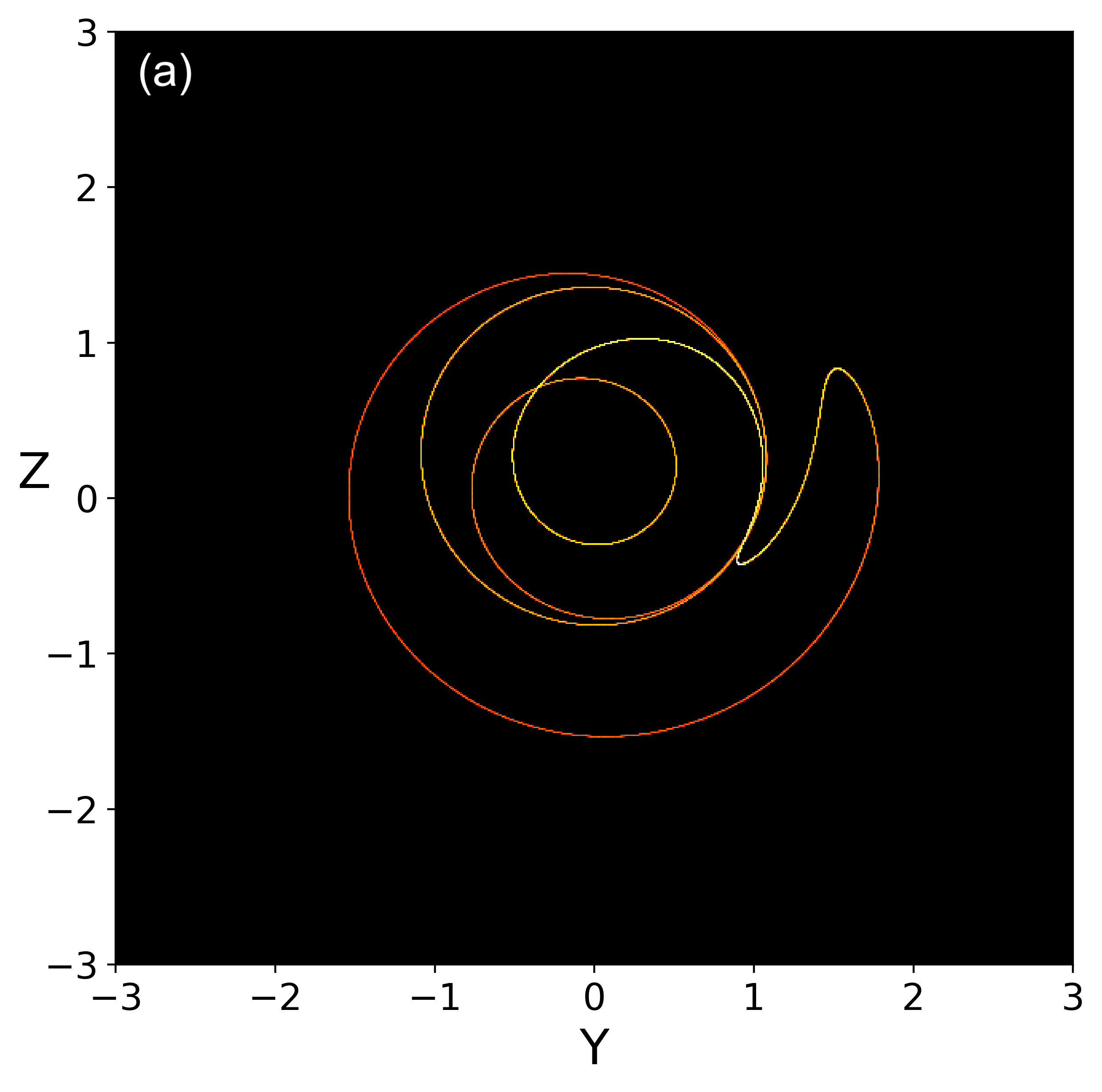}
  \end{minipage}%
  \begin{minipage}{0.45\textwidth}
    \centering
    \includegraphics[width=\textwidth]{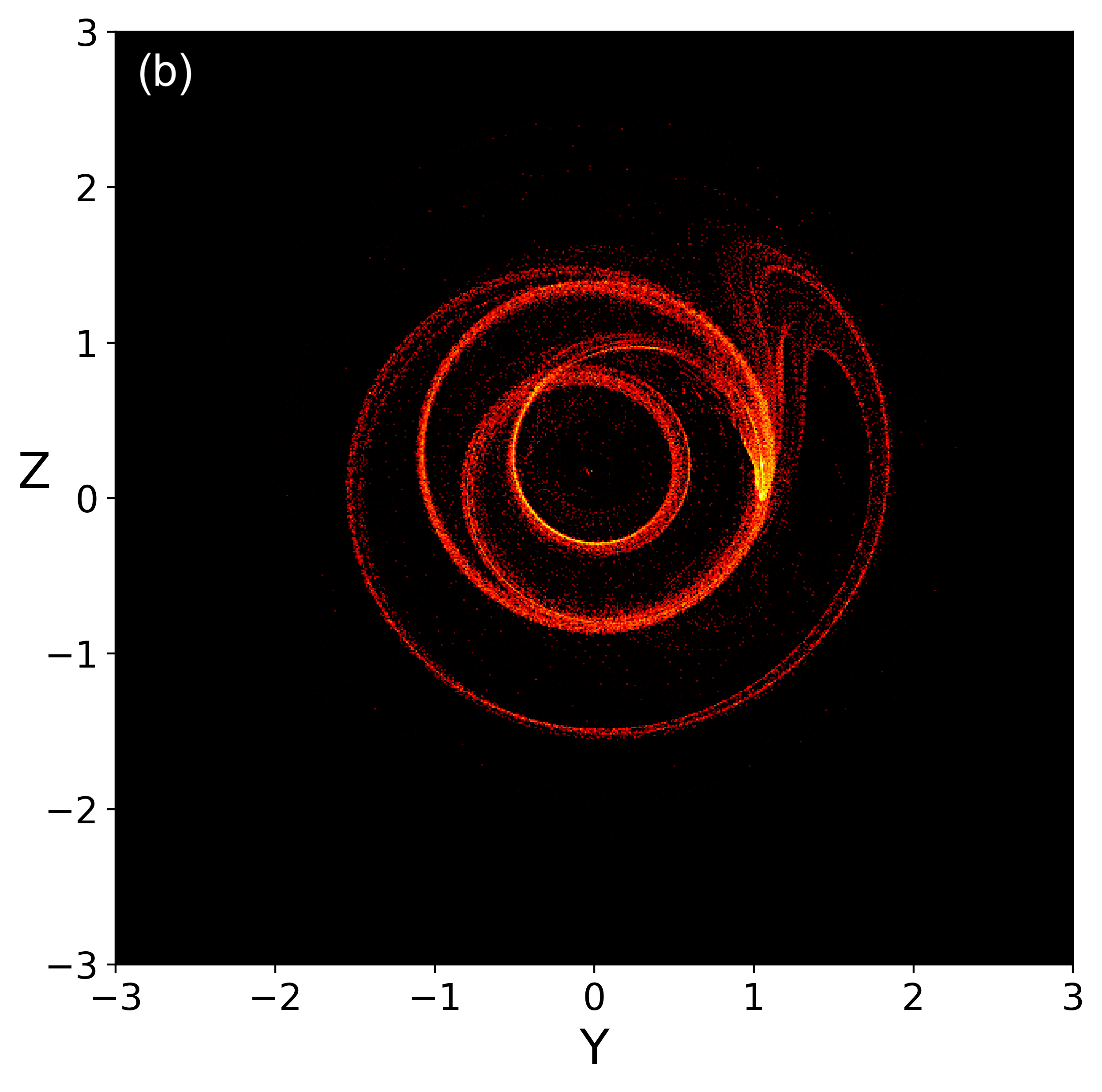}
  \end{minipage}
   \caption{Summer heat map (a) of the forward attractor for $F \equiv 6$; and (b) of the snapshot attractor at the time ${t}_1=48.6$~time units into the year, when $F{t}_1) = 6$. The heat maps in panels (a) and (b) were made using $5 \cdot 10^4$ points that are the intercepts at $t = t_1$ of the trajectories in the ensemble. The initial conditions for the ensemble were chosen at random in the same cube $D \subset \mathbb R^3$ as in Fig.~\ref{fig:colonna}. To capture the two attractors sufficiently accurately, we had to let the ensemble evolve for a time larger than the convergence time of $\tau_c \simeq 5$ years.}
   \label{fig:PBA_F6}
\end{figure*}
\begin{figure*}
  \begin{minipage}{0.45\textwidth}
    \centering
    \includegraphics[width=\textwidth]{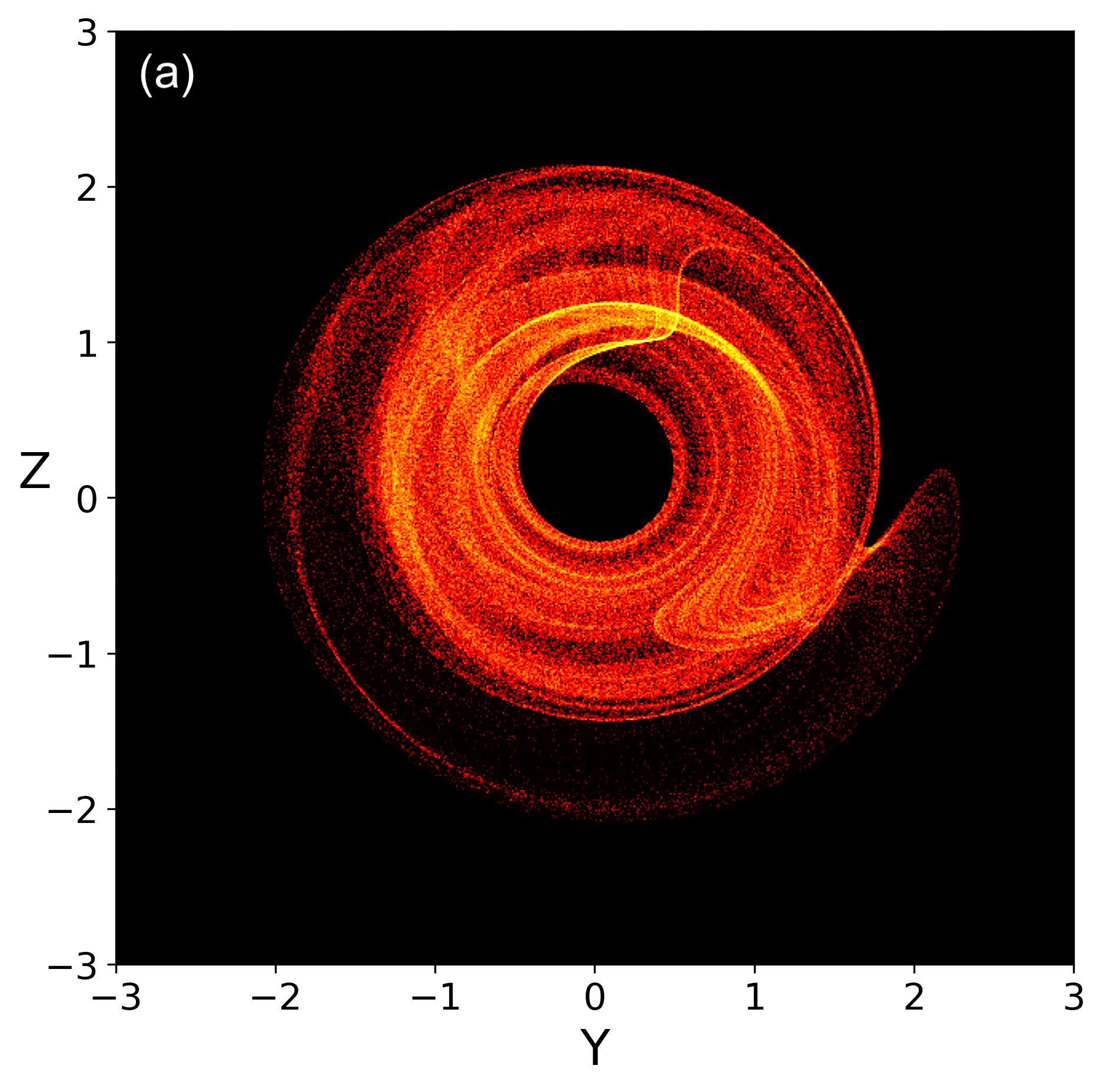}
  \end{minipage}%
  \begin{minipage}{0.45\textwidth}
    \centering
    \includegraphics[width=\textwidth]{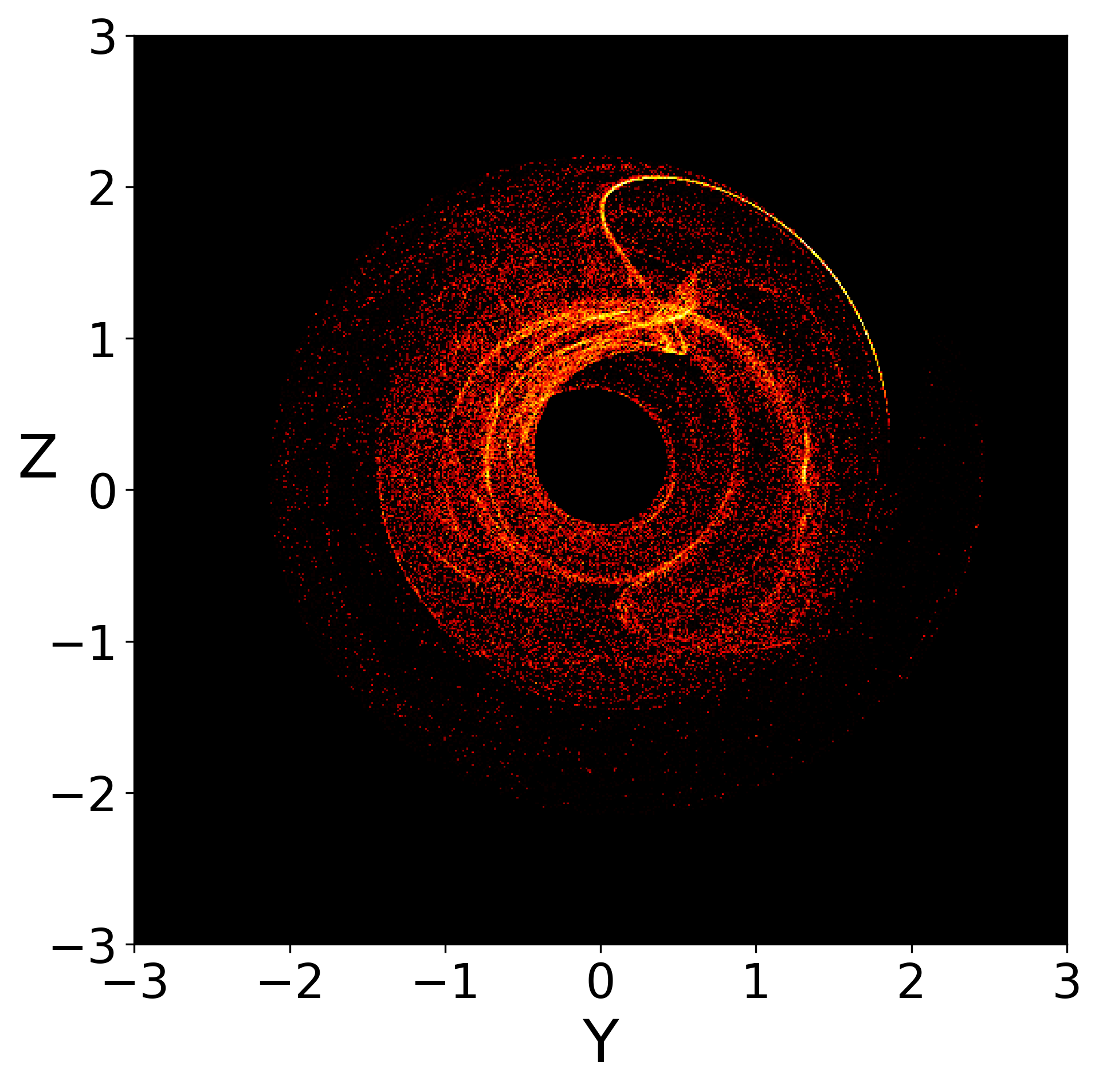}
    
  \end{minipage}
\caption{Same as Fig.~\ref{fig:PBA_F6}  but for the winter season, (a) with $F \equiv 8$; and (b) with ${t}_1=12$~time units, when $F({t}_1)=8$. }
  \label{fig:PBA_F8} 
\end{figure*}
The snapshot attractor contains all the information needed to study the ``climate'' of the model, and it suffices for our purposes herein to inspect its projection onto the $(Y,Z)$-plane for a comparison between different cases. This projection is plotted in the following figures as a {\em heat map} or two-dimensional histogram \citep{Wilkinson.2009, keno} that records the number of points of the attractor that fall within a pixel on this plane. To be precise, the heat map is built as a two-dimensional histogram over the square $\{-3 \le  Y \le 3, -3 \le Z \le 3\}$, with 500 or 600 bins in each direction, i.e., a total of $2.5 \cdot 10^5$ or $3.6 \cdot 10^5$ pixels; when tested, the difference in resolution does not seem to make much of a difference in the resulting histogram. \\
An alternative analysis could be carried out by determining the Poincaré maps that arise from the intersection of the trajectories on the attractor with a well chosen plane in the model's phase space. The choice of visualizing the invariant measure on the attractor as the heat map of a planar projection follows an example in \cite{keno}. The projection of the attractor on the $(Y,Z)$-plane is stroboscopic when only seasonal forcing acts on the system, given the periodic recurrence of the same pattern. In practice, the heat maps of projections onto the $(Y, Z)$-plane herein are obtained simply by counting all the points with given $(Y, Z)$ coordinates at a prescribed epoch or within a prescribed time interval, as described in the caption of the figure of interest. \\

\subsection{Seasonal forcing}
\label{sec:seasonal}
Recall that, in the autonomous case, the L84 model exhibits either a periodic or a chaotic behavior, for the perpetual summer forcing of $F \equiv 6$ or winter forcing of $F \equiv 8$, respectively. When a time-dependent seasonal forcing is included,  this feature becomes less obvious. The picture changes for the summer, cf. Fig.~\ref{fig:PBA_F6}, while it remains generally true for the winter season, cf. Fig.~\ref{fig:PBA_F8}.\\
During perpetual summer ($F \equiv 6$), the  heat map in Fig.~\ref{fig:PBA_F6}(a) corresponds simply to a projection of the two limit cycles in Fig.~\ref{fig:basins}(a) onto the $(Y,Z)$ plane and it is clearly one-dimensional, i.e, a simple line. In the presence of seasonal forcing, though, the chaotic character of the winters "thickens" the heat map during the summers, too, since the model behavior at the times $t_1$ and nearby becomes chaotic as well, as first observed by \cite{lorenz90} and illustrated in our Fig.~\ref{fig:PBA_F6}(b). In simple meteorological terms, chaotic winter seasons disrupt the more regular summer circulation and induce less regular variability in the summers as well. Thus, periodicity and regularity are lost, and higher dependence on initial conditions leads to lower predictability on S2S time scales during summers, too. \\
When $F=8$, although the behavior is still chaotic in both panels of Fig.~\ref{fig:PBA_F8}, there is a clear change in the shape of the attractor from the autonomous case in panel (a) to the nonautonomous one in panel (b). The autonomous case shows greater uniformity in the high-intensity area of the heat map, with a near symmetry between the two components, $Y$ and $Z,$ of the eddies. Overall, the area of the heat map shrinks in the nonautonomous case, while a region with significantly higher values of $Z$ is visited by the system, which appears as the brighter fiber in Fig.~\ref{fig:PBA_F8}(b), and corresponds to a stronger activity of the $Z$-wave component. Although it is bounded, the energy of the L84 model can change over time and it is not conserved in the nonautonomous case. Here, the wave contribution increases and extreme values are more frequently assumed.

\begin{figure*}
  \begin{minipage}{0.45\textwidth}
    \centering
    \includegraphics[width=\textwidth]{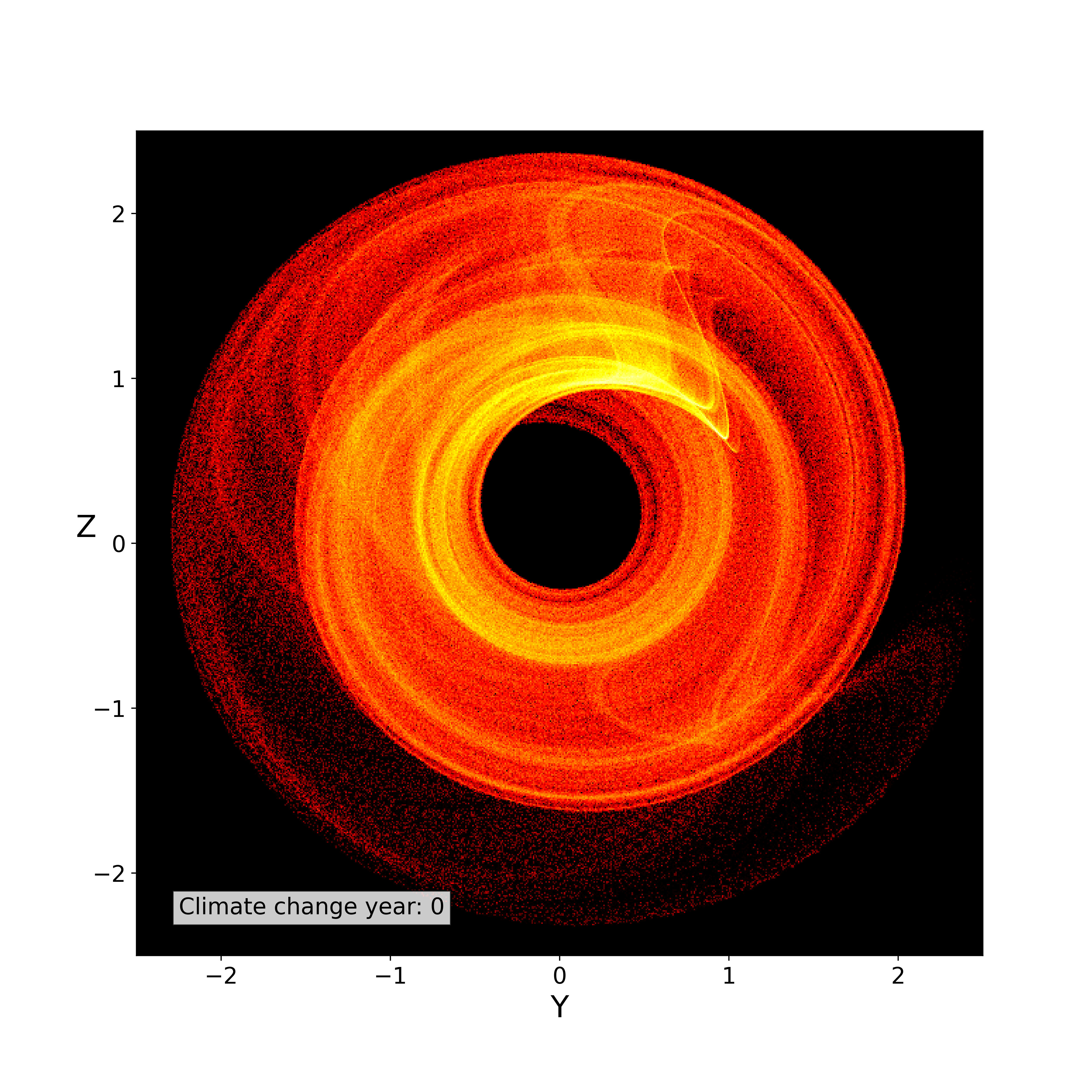}
  \end{minipage}%
  \begin{minipage}{0.45\textwidth}
    \centering
    \includegraphics[width=\textwidth]{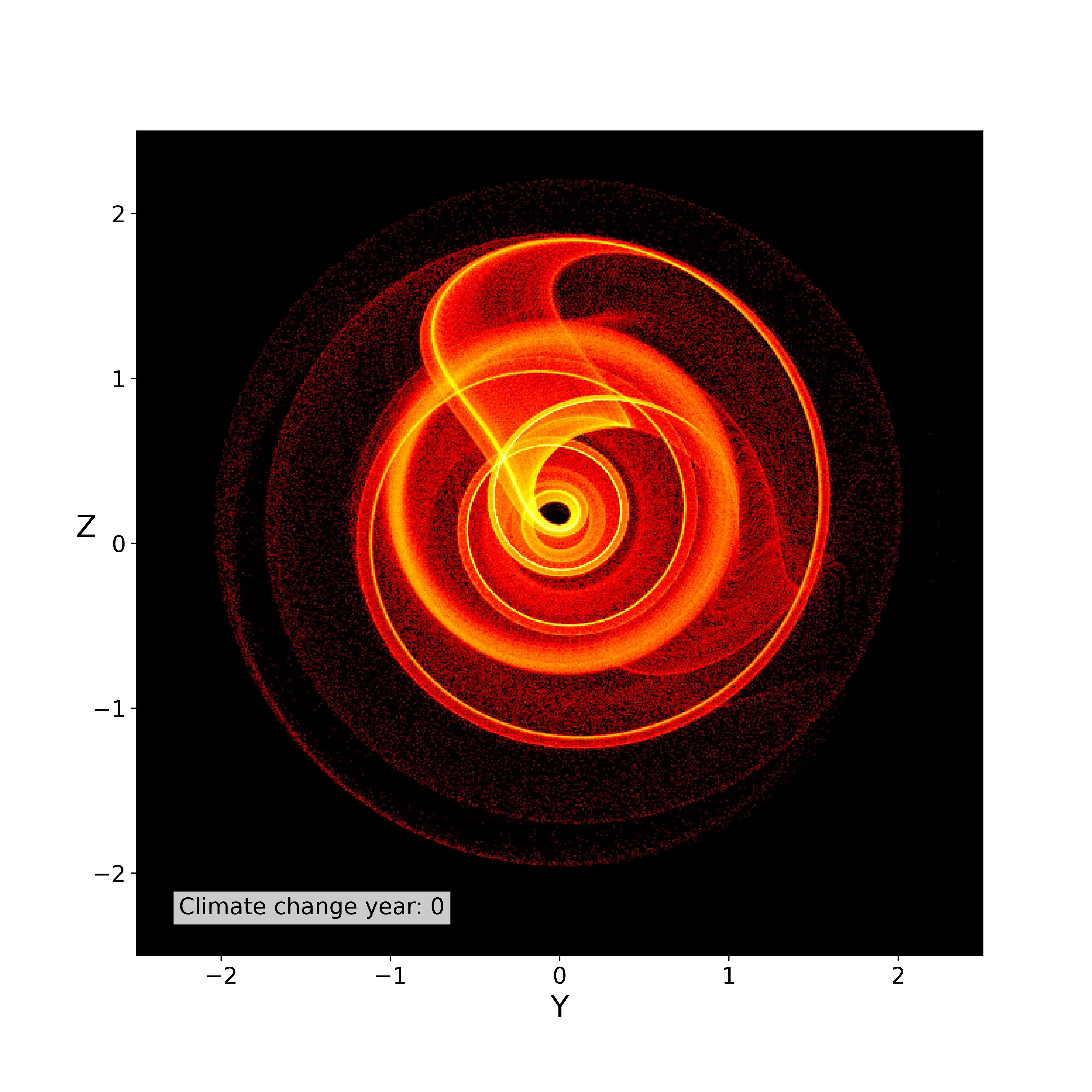}
    
  \end{minipage}
\caption{Projection of the attractor on the $(Y,Z)$-plane for the months of January and July,
 		before climate change. }
  \label{fig:Jan+July} 
\end{figure*}


\subsection{Climatic trends}
We next used the L84 model to study both qualitatively  and quantitatively the effects of climatic trends in the meridional heat contrast on the mid-latitudes' atmospheric circulation. This aspect is the fundamentally novel one of our L84 model investigation. Both periodic and monotonic forcing were considered and their combined effect was analyzed. Moreover, either a positive or a negative linear trend were applied to the forcing $F(t)$ in order to capture the effect of climate change on the lower and upper levels of the atmosphere, respectively.

Generally, by applying an aperiodic forcing to the system, one expects the attractor to deviate from a stationary or periodic state and to change shape in time. The visual changes in the attractor were also analyzed quantitatively by determining the change in the first four moments of the distributions of the prognostic variables $X, Y$ and $Z$, namely their mean, variance, skewness and kurtosis. \\
The full months of January and July are taken as references for the effects of climate trends. One could also consider longer intervals, like a full three months for summer and winter, to monitor anthropogenic effects on climate. Whether summer or winter, the attractor described in this subsection is a collection of the snapshots of every moment $t$ for either interval, be it one month or longer. \\
Figures~\ref{fig:Jan+July}(a,b) show the attractor for January and July, respectively, in the absence of any trend in the forcing. 
The heat map in panel (a) illustrates chaotic behavior that is in accordance with the choice of the parameter range for the winter season, while panel (b) is more regular, as expected for the summer season. Still,  chaotic features — indicated by a more diffuse heat map in Fig.~\ref{fig:Jan+July}(b) than in Fig.~\ref{fig:PBA_F6}(b), i.e., in monthly mean map than in a single snapshot — are visible, too, as expected from the perturbing effects of the winter circulation on that of the summers, cf.~\cite{lorenz90} and  section~\ref{sec:seasonal} herein. \\ 
As stated above, the forcing $F(t)$ here follows equation \eqref{F(t)}, with no periodic component. To properly capture the effects of the trend on the attractor, the trendless heat maps in Fig.~\ref{fig:Jan+July} are visually compared with three distinct years within the interval wherein the trend is active, specifically the $5^{\rm{th}}$, $50^{\rm{th}}$, and $100^{\rm{th}}$ year of climate change. \\
Figures~\ref{fig:january_negative} and \ref{fig:july_negative} show the changes in the winter and the summer attractor, respectively, during 100 years of a climate change interval in the forcing. In each figure, panels (a--c) follow the negative trend, while panels (d--f) correspond to a positive one. The slope of the trend is $\alpha = 2/100$~year$^{-1}$, so that $F$ decreases or increases by 2 nondimensional units per century.\\
The attractor is affected but little by the forcing 5 years after the trend has started, in all four cases, whether winter or summer and for a trend that is negative or positive; see Figs.~\ref{fig:january_negative}(a,b) and \ref{fig:july_negative}(a, b), respectively. Substantial changes, though, appear after 50 years (panels (b, e)) and especially after 100 years ((panels (c, f))).
By the end of the $100^{\rm{th}}$ year, the L84 model displays a much more regular behavior for the month of January, as well as July, in three of the four cases, the exception being July with a positive $F$-trend, i.e., Fig.~\ref{fig:july_negative}(f). In the latter case, the waves $(Y,Z)$ are both more vigorous and more irregular than at present, before the positive forcing trend started. To the contrary, for summer and a negative trend, the behavior reduces to a periodic wave propagation with a very small amplitude; see Fig.~\ref{fig:july_negative}(c). \\   
The quantitative changes of the distribution of the westerly wind $X$ and the wave energy $E_{YZ}  = (Y^2+Z^2)/2$ are shown in Fig.~\ref{fig:statistics}, which reports the dependence of the first four statistical moments of these two variables on time, for January and for July, including both a negative and a positive trend.
As far as the mean wind velocity is concerned (black curves), no major change in the moments is observed for the month of January all the way to the end of the century (Figs.~\ref{fig:statistics}(a,c)), while in July for the negative trend there is a sharp drop at roughly 50 years and strong fluctuations afterwards, with a return to a smooth evolution towards the end of the interval (Fig.~\ref{fig:statistics}(e)). \\ 
The variances of the mean and wave components (yellow curves) are both quite flat, in summer and in winter, except for a smooth decrease in the first half-century for the wave component in summer, given a negative climate trend (panel (f)), and a sharp decrease in the last decade for the mean wind, for the same situation (panel (e)). \\ 
The skewness of both distributions (blue curves) does not exhibit considerable changes during the whole century of climate trends, except a smooth decline along with the variance in the same time interval and situation as the variance, i.e., negative trend for July (panel (e)). Thus, the skewness is mostly close to 0, meaning that the distribution is fairly symmetric. This indicates that the energy tends to be equally distributed between the two wave components, as expected from the L84 model's symmetry in $Y$ and $Z$; see comment in Sec.~\ref{ssec:auto}. \\
The kurtosis of the distribution of the wave energy $E_{YZ}$ varies substantially in both winter and summer, as well as for both negative and positive trends. These variations are quite distinct from each other: uniform decrease in winter for a negative trend (panels (a,b)); sharp drop at roughly 60 years, followed by continued decrease to the end of the century in winter for $E_{YZ}$ and a  positive trend; smooth decrease in the first half century and flat to the end for the wave energy in summer with negative trend (panel (f)); and decrease during roughly 60 years followed by a sharper increase in panel (f). \\
This variability is quite interesting, since  large kurtosis means the distribution is heavy tailed and small kurtosis means that the opposite holds. Thus the distribution of extremes tends to change, and change in different ways, for summer vs. winter and for distinct trends in the anthropogenic forcing we are simulating.\\
The evidence illustrated in Figs.~\ref{fig:january_negative}--\ref{fig:statistics} is at variance with a simple, direct relation between the thermal response of the mid-latitude atmosphere to climate change and its effects on its mean intensity, its eddy activity and its extremes. We will return to a discussion of these complexities in jet and wave changes in Sec.~\ref{sec:concl}.  

\subsection{An extreme effect of time-dependent forcing}
\label{sec:trend_ultima}
We now illustrate a case in which the time-dependent forcing that includes both a seasonal and a trend component gives a particularly striking result. Figure~\ref{fig:PBA_F199}(a) shows the projection of the autonomous forward attractor onto the $(Y,Z)$-plane for an ensemble of initial conditions that were forced with a value of $F \equiv 1.99$ that happens to correspond to the forcing of the system at the beginning of July of year 150 of climate change with a negative trend. \\
This heat map shows that, for $F = 1.99$, the system exhibits two separate attractors, namely a limit cycle (red circle in the figure) and a fixed point $(Y^*, Z^*)$. In this case, with the ensemble initialized in the same cube $D \subset \mathbb R$ as before, around $98~\%$ of the trajectories are attracted to the fixed point, to which the white arrow points. \\
Figure~\ref{fig:PBA_F199}(b) shows the snapshot of the month of July for year 150 after climate change began. Here, the fixed point at $(Y^*, Z^*)$ has disappeared, while the limit cycle has thickened since it attracts all the orbits over the entire month of July. \\
Since nature does not reset itself, we are not given the chance to observe the state of the system at the same epoch $t$ multiple times. On the other hand, continuous time-dependent effects on the system's evolution do occur. While the forward attractor of the autonomous system only reproduces the asymptotic state for perpetual, constant forcing, the snapshot attractor allows us to observe the effects of the past evolution on the present state, including forcing effects that change with time.

\section{Conclusions and final remarks}
\label{sec:concl}

The conceptual model of the mid-latitude atmospheric circulation of \cite{lorenz84,lorenz90} has already made important contributions to the applications of dynamical systems theory to the climate sciences \citep{shilnikov, vanveen2001, Broer, freire2008multistability}, as well as to additional areas \citep{mangiarotti}. Herein, we have focused on the way in which this model can help to better understand the effects of time-dependent seasonal forcing and anthropogenic forcing trends on climatic systems with intrinsic variability \citep{Ghil.Lucar.2020}. \\
To start, we have applied bifurcation analysis to the time-independent, autonomous model and compared our results with those of previous work. Using both the cross-latitude forcing $F$ and the damping parameter $a$ as control parameters, we found similar results in both cases: the system undergoes first a double-fold bifurcation that leads from a single to two stable steady states, separated by an unstable one, and then a Hopf bifurcation giving rise to a stable limit cycle; see Figs.~\ref{fig:polynomial}--\ref{fig:bif_kit}. Coexistence of a stable fixed point with such a cycle is also possible, as are more complicated modes of behavior that follow (Figs.~\ref{fig:bifurcation_a} and \ref{fig:colonna}). \\
For perpetual summer-level forcing of $F=6,$ we did confirm the crucial finding of \cite{lorenz90} that two oscillatory solutions coexist (Fig.~\ref{fig:basins}(a)), one with a period of 7 days, roughly comparable to that of the life cycle of baroclinic instabilities, the other with a period of 35 days, roughly comparable with that of mid-latitude intraseasonal oscillations \citep{ghil_S2S}. Using advanced tools from the open-access Julia ecosystem \citep{Datseris2018,Datseris.2022}, we were able to delineate the attractor basin boundaries of the two corresponding limit cycles, which appear to be fractal (Fig.~\ref{fig:basins}(b)).

\begin{figure*}[!ht]
	\includegraphics[width=0.9\textwidth]{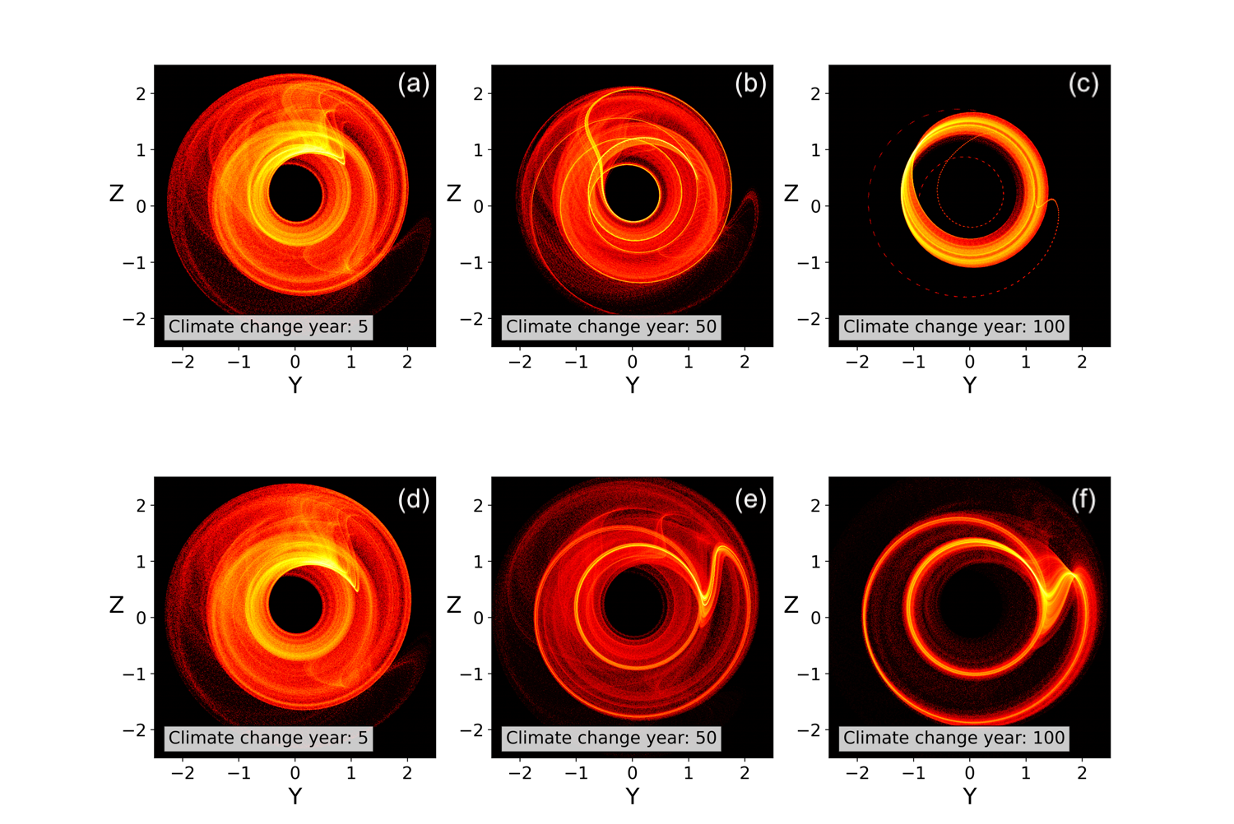}
	\caption{Projection of the attractor on the $(Y,Z)$-plane for the month of January of years (a,d) 5, (b,e) 50 and (c,f) 100 during a climate trend that is negative in panels (a--c), and positive in panels (d--f); the trend has the slopes $\alpha = \mp 2/100$~year$^{-1}$, respectively}
	\label{fig:january_negative}
\end{figure*}
\begin{figure*}[!hb]
	\includegraphics[width=0.9\textwidth]{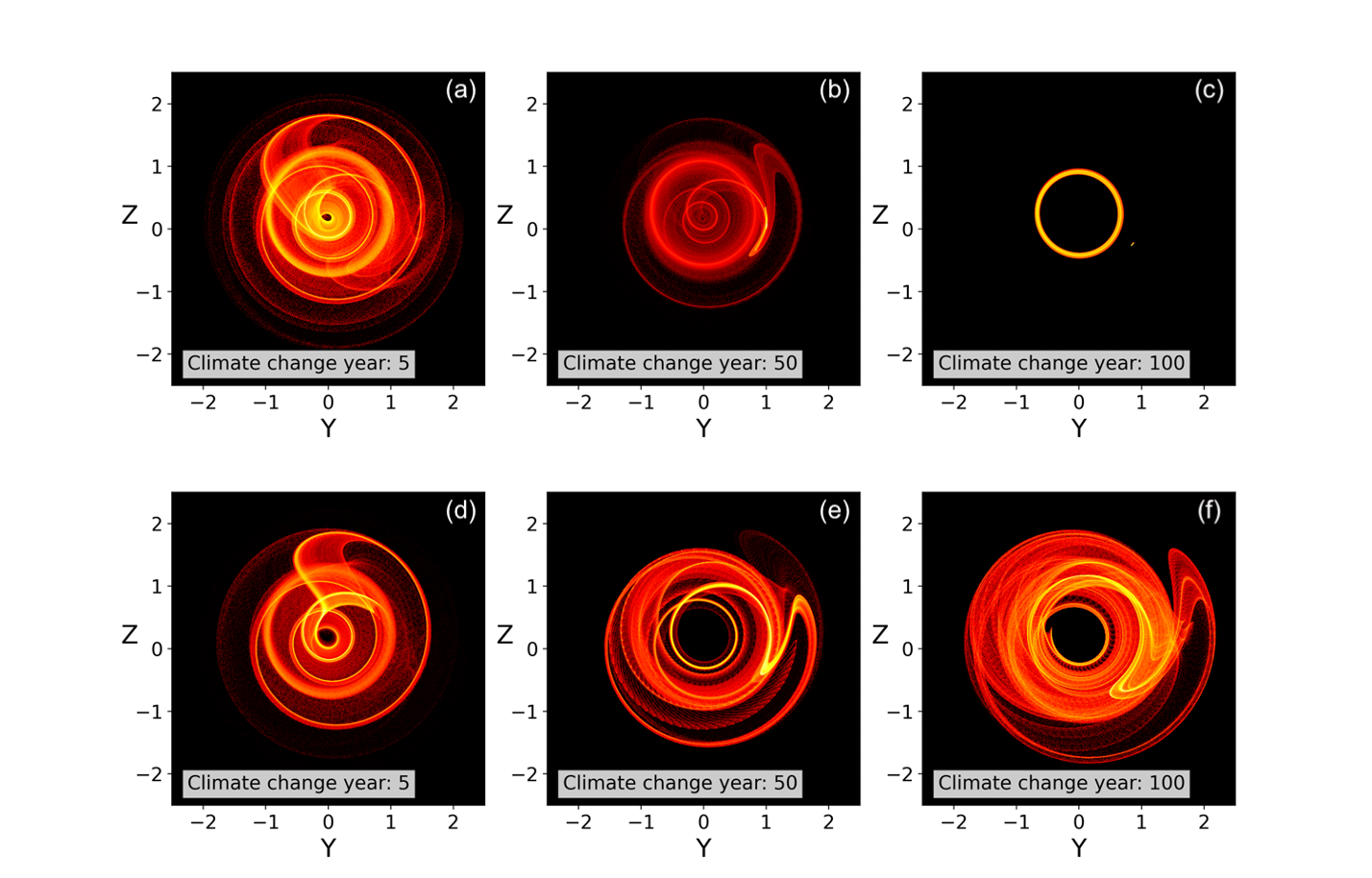}
	\caption{Same as Fig.~\ref{fig:january_negative}, but for the month of July.}
	\label{fig:july_negative}	
\end{figure*}

\clearpage

\begin{figure*}
	\begin{minipage}{0.45\textwidth}
		\centering
		\includegraphics[width=\textwidth]{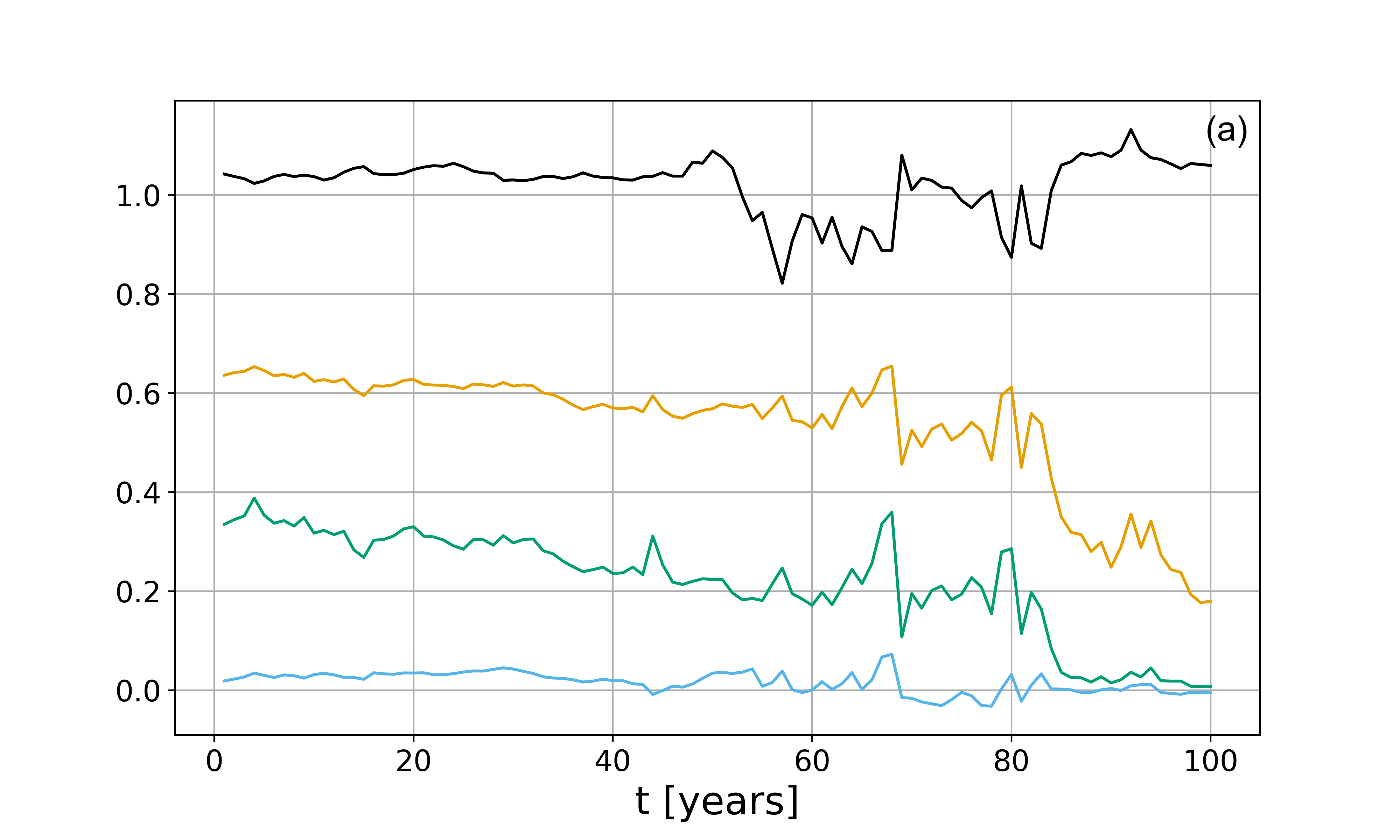}
	\end{minipage}%
	\begin{minipage}{0.45\textwidth}
		\centering
		\includegraphics[width=\textwidth]{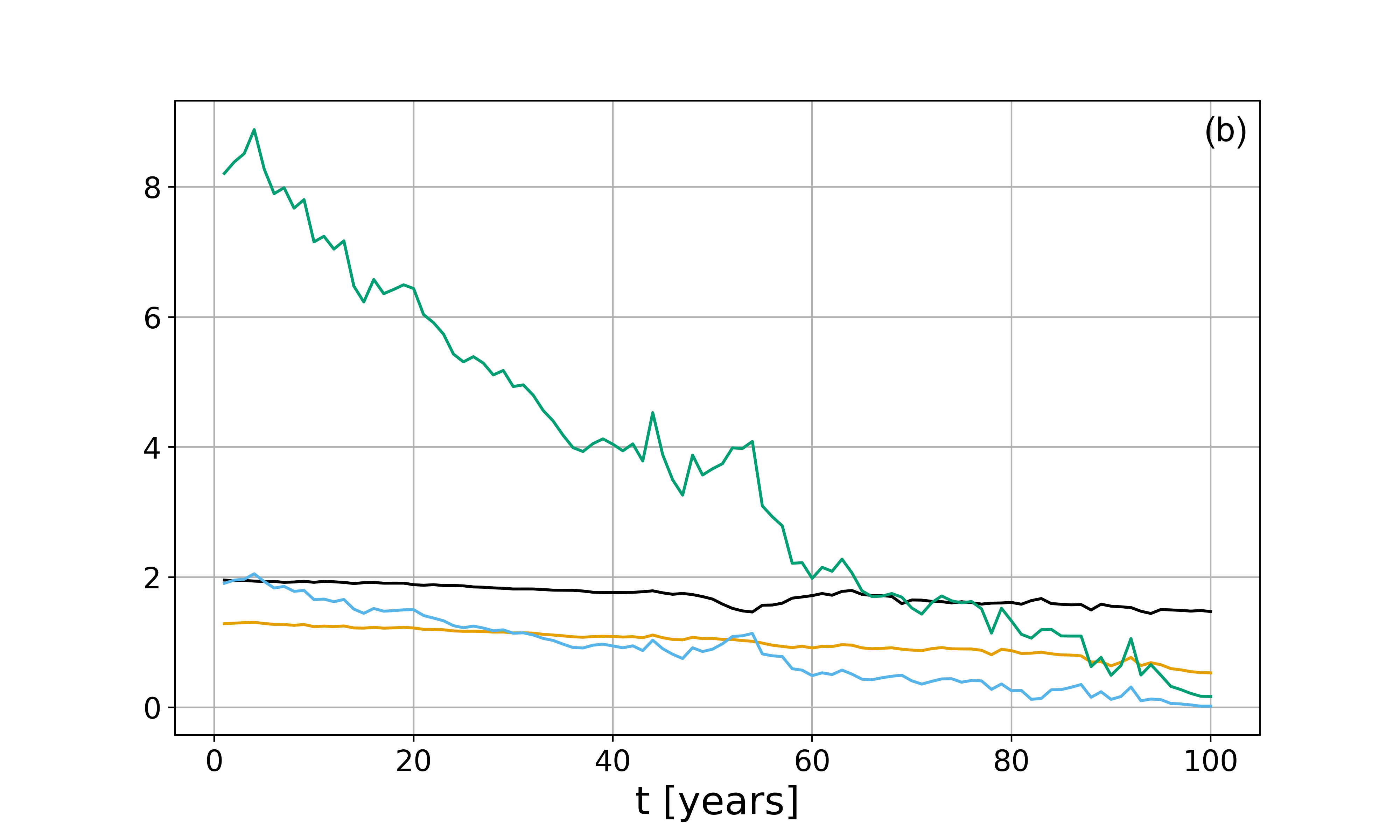}
	\end{minipage}
	
	\label{} 
	
\end{figure*}

\begin{figure*}
	\begin{minipage}{0.45\textwidth}
		\centering
		\includegraphics[width=\textwidth]{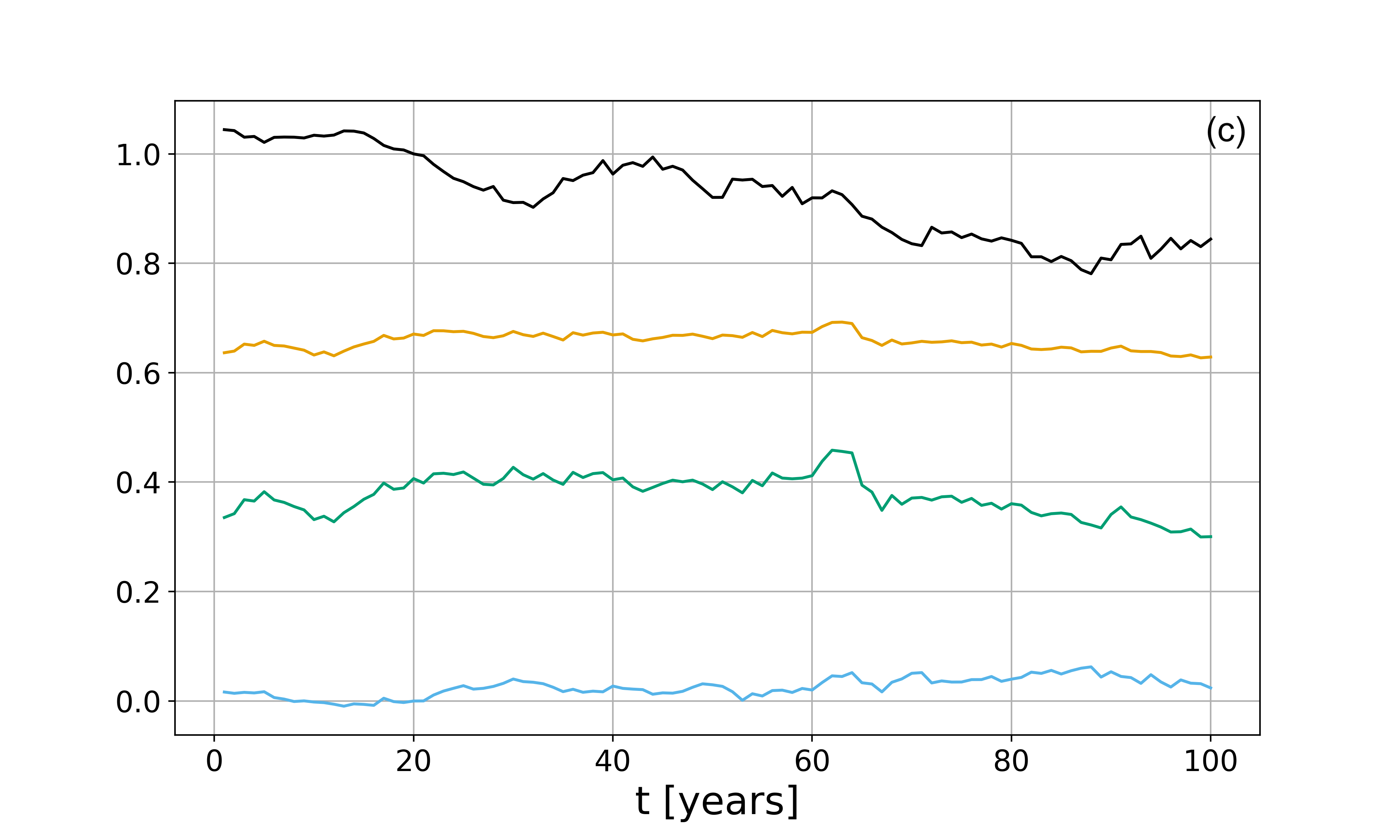 }
	\end{minipage}%
	\begin{minipage}{0.45\textwidth}
		\centering
		\includegraphics[width=\textwidth]{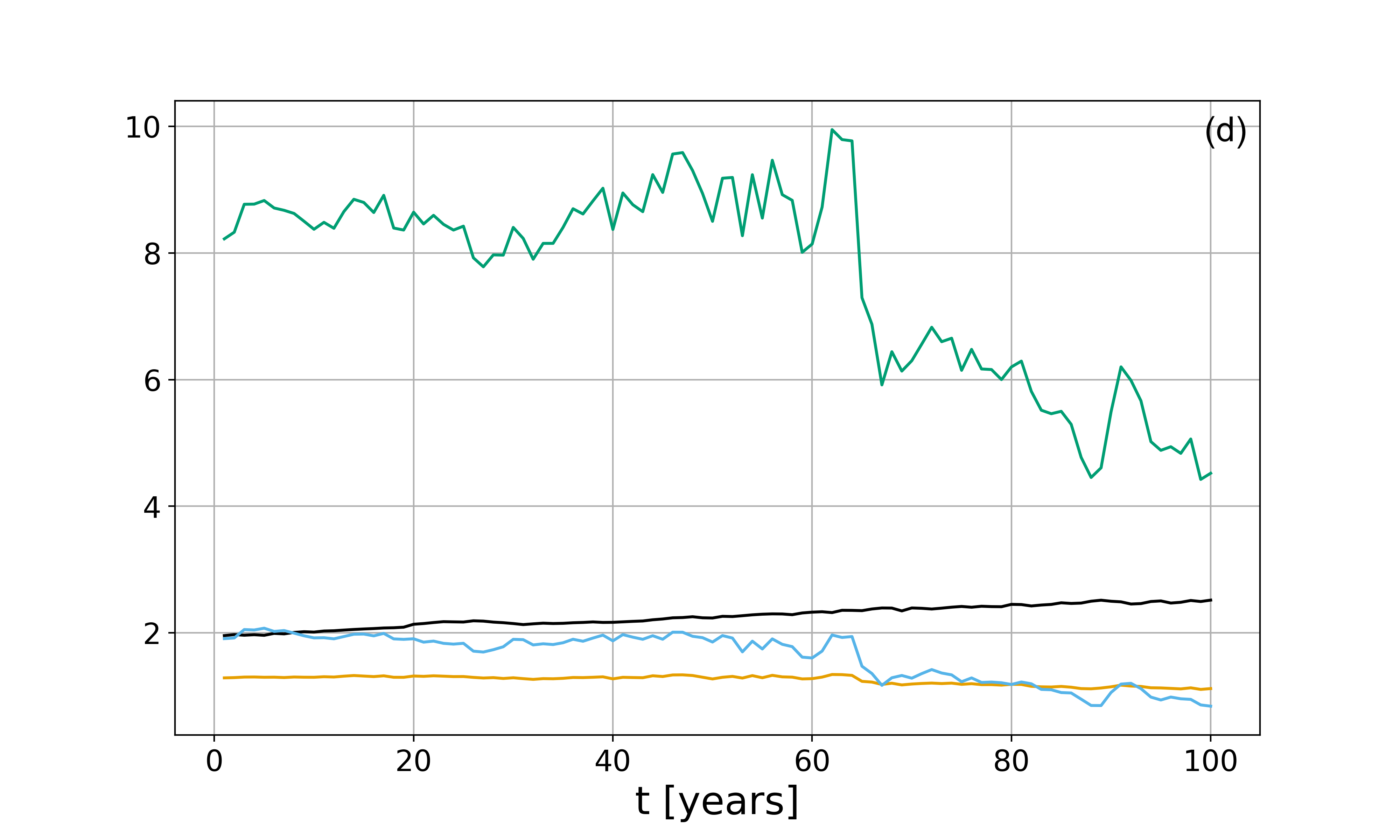}
	\end{minipage}
	
	\label{} 
	
\end{figure*}

\begin{figure*}
	\begin{minipage}{0.48\textwidth}
		\centering
		\includegraphics[width=\textwidth]{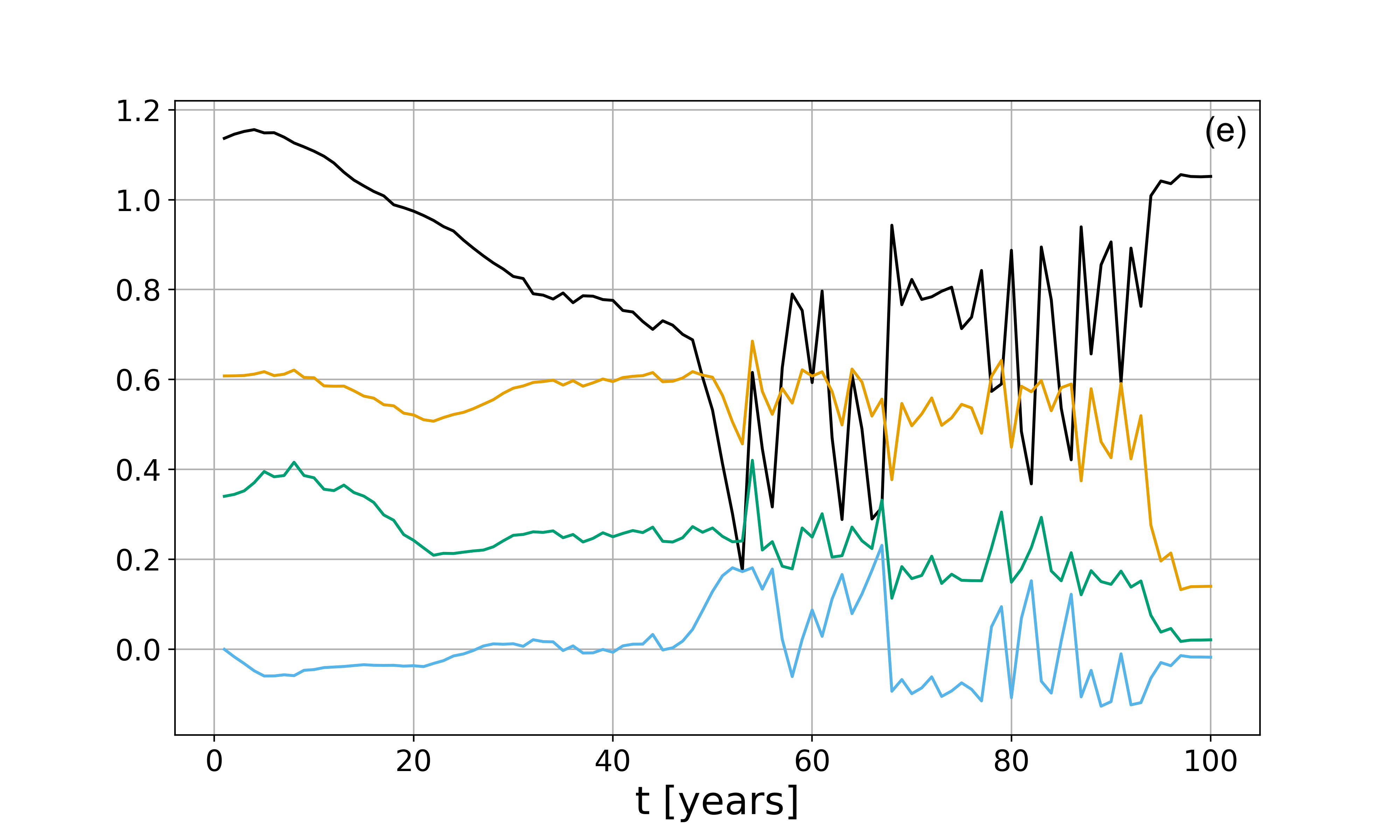}
	\end{minipage}%
	\begin{minipage}{0.48\textwidth}
		\centering
		\includegraphics[width=\textwidth]{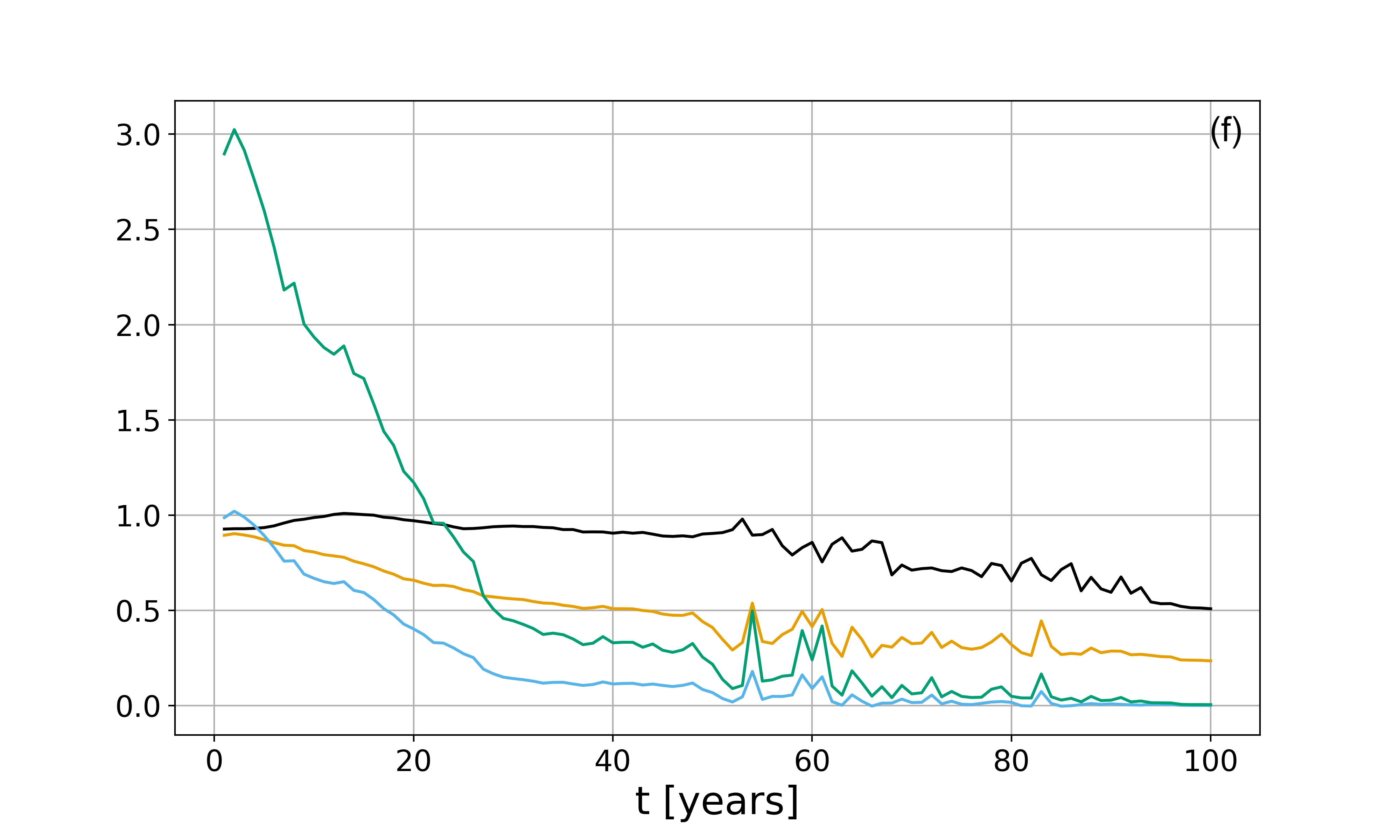}
	\end{minipage}
	
	\label{} 
	
\end{figure*}

\begin{figure*}
	\begin{minipage}{0.48\textwidth}
		\centering
		\includegraphics[width=\textwidth]{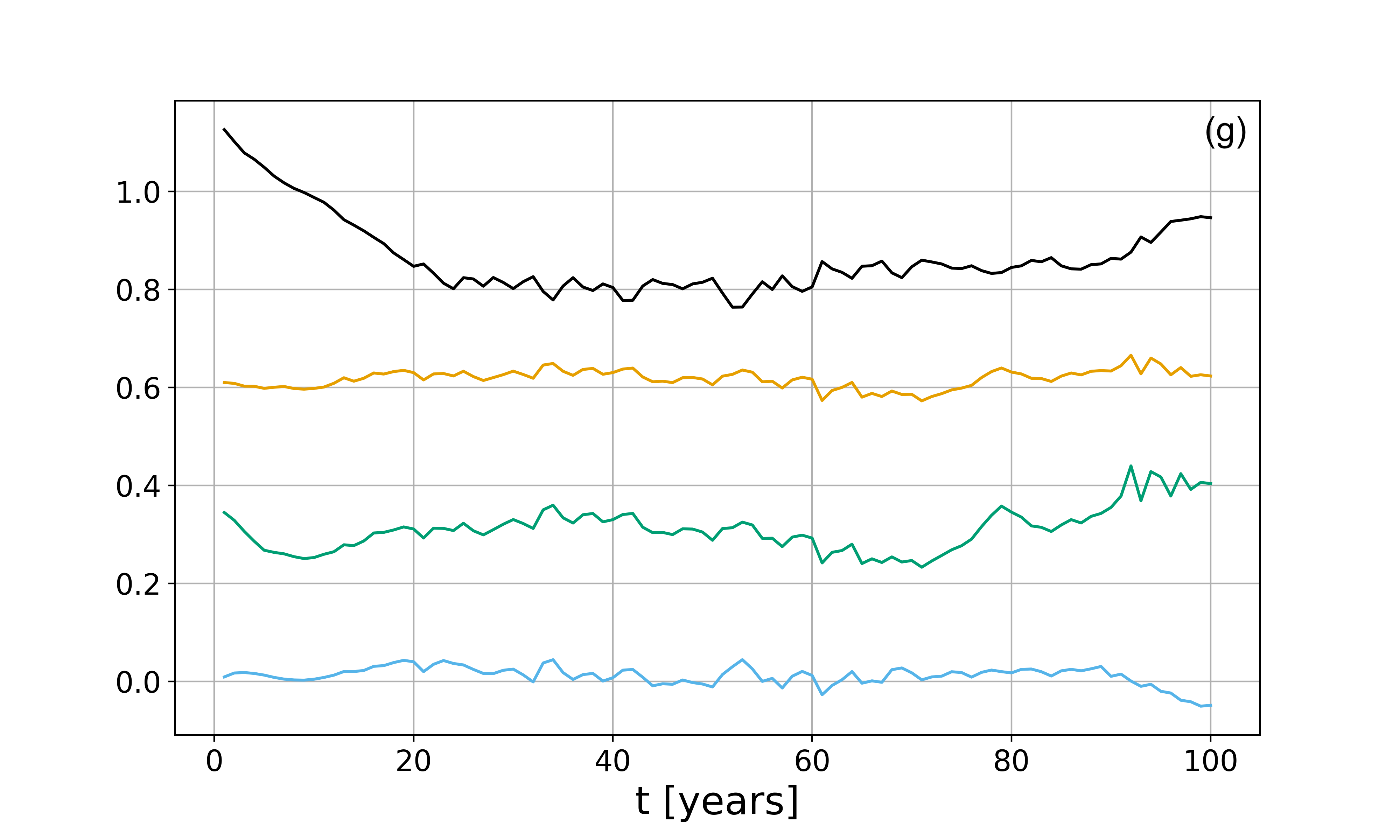 }
	\end{minipage}%
	\begin{minipage}{0.48\textwidth}
		\centering
		\includegraphics[width=\textwidth]{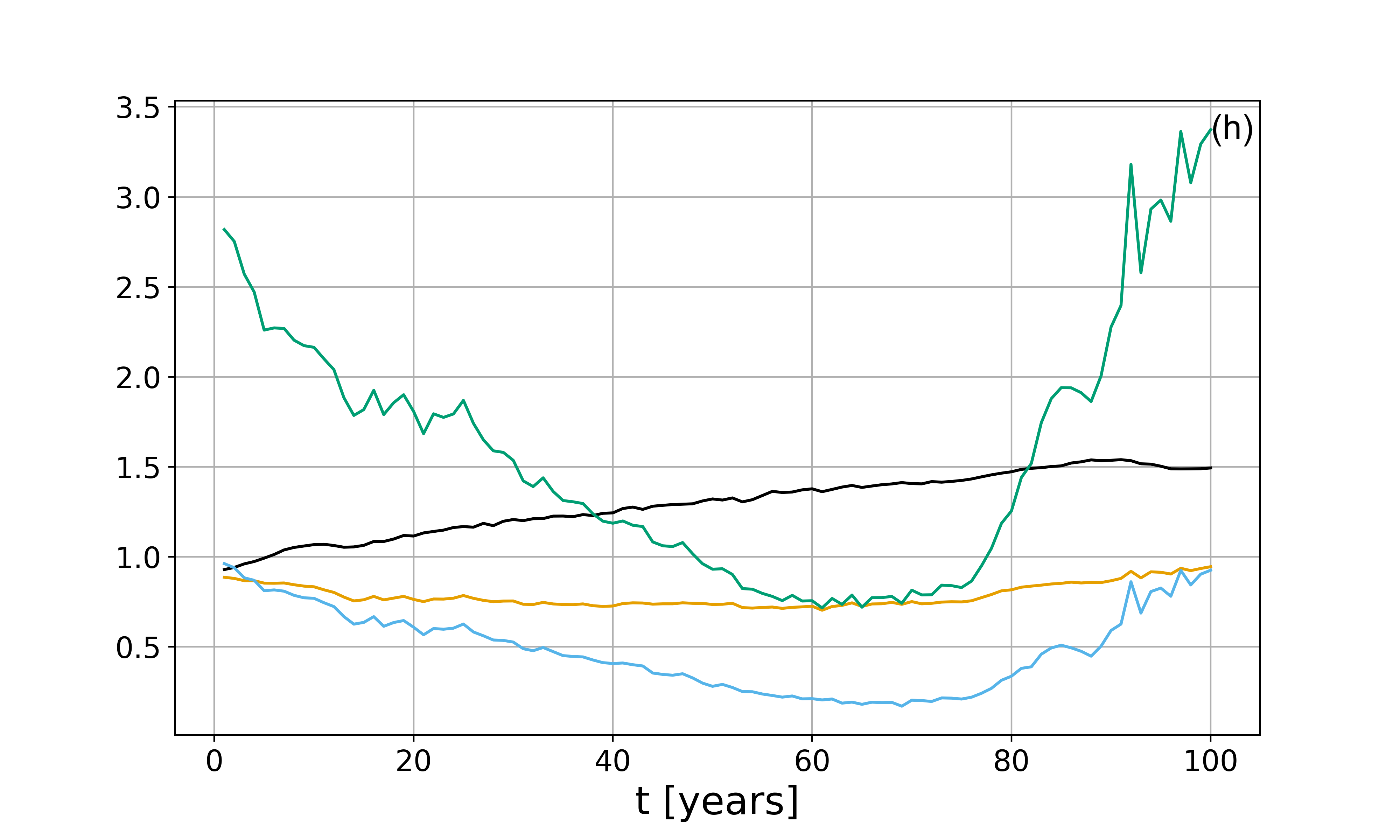}
	\end{minipage}
	\caption{Statistical moments on the attractor over time for (a,c,e,g) the westerly flow $X$ (left column); and for (b, d, f, h) the wave energy $E_{Y, Z} = Y^2+Z^2$ (right column). Rows are: (a,b) January, negative trend; (c,d) January, positive trend; (e,f) July, negative trend; and (g,h) July, positive trend. Legend: black (mean), yellow (variance), blue (skewness), green (kurtosis). All vertical axes are in nondimensional units.}
	\label{fig:statistics} 
	
\end{figure*}

\begin{figure*}
	\begin{minipage}{0.45\textwidth}
		\centering
		\includegraphics[width=\textwidth]{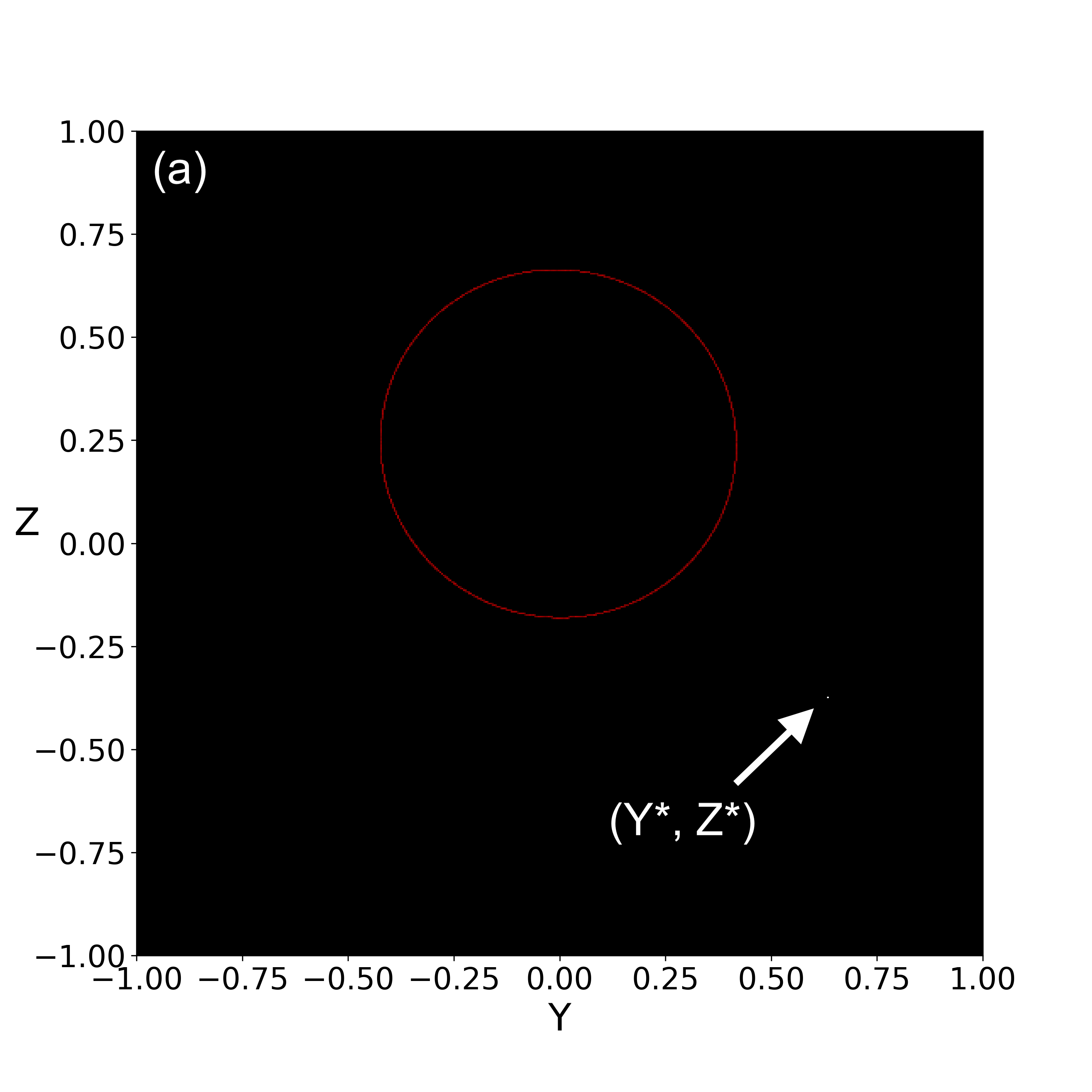}
	\end{minipage}%
	\begin{minipage}{0.45\textwidth}
		\centering
		\includegraphics[width=\textwidth]{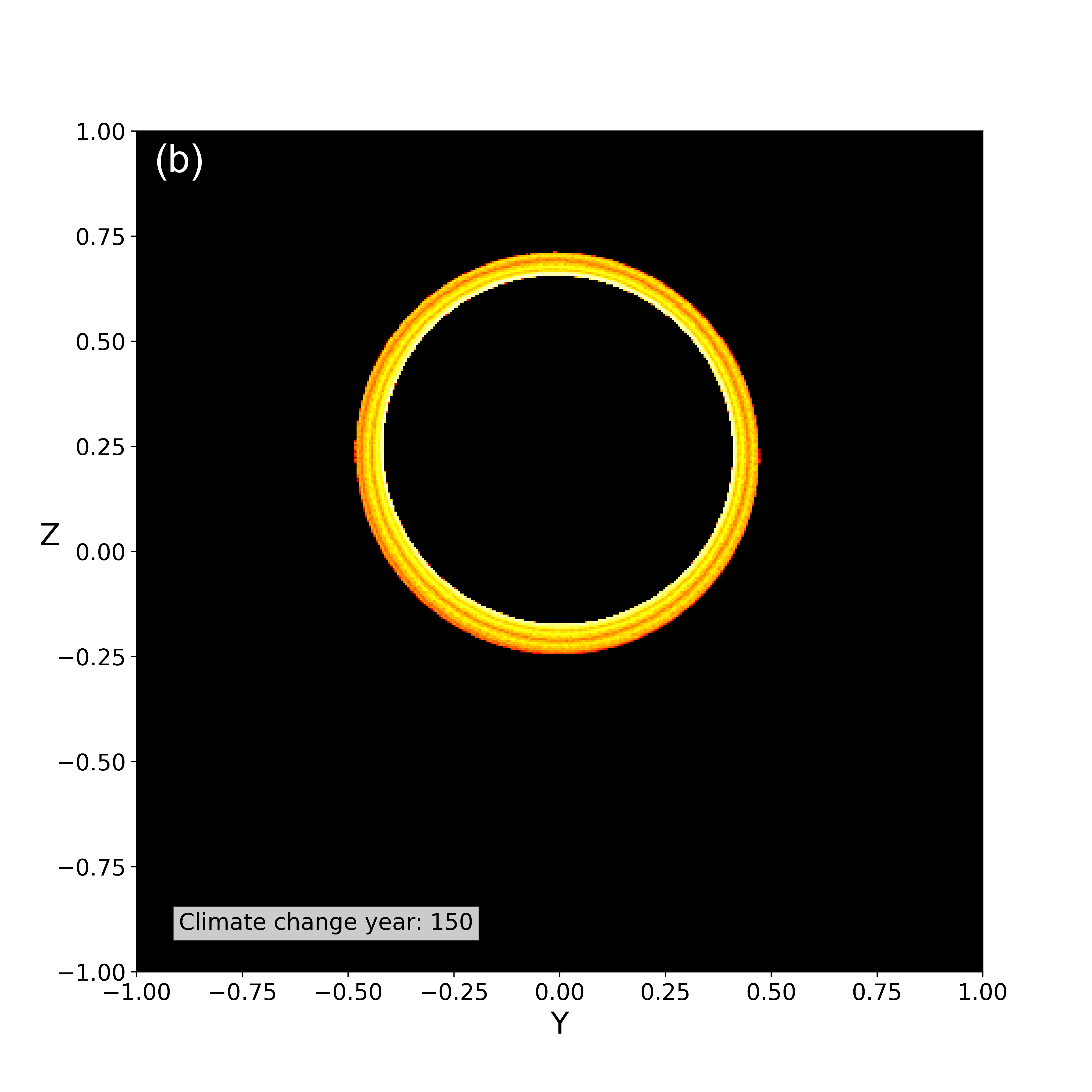}
	\end{minipage}
	\caption{Projection of the attractors on the $(Y, Z)$-plane in the case of a combined seasonal and trend forcing: (a) forward attractor for a perpetual summer at $F \equiv 1.99$; and (b) snapshot attractor for the month of July at year 150 of climate change with a trend of $\alpha = - 2/100$~year$^{-1}$ in the meridional heat contrast; the instantaneous thermal forcing in panel (b) also equals $F = 1.99$ at the beginning of the month. }
	\label{fig:PBA_F199} 
\end{figure*}
Proceeding to the time-dependent, nonautonomous case that really motivated this study, we introduced first the various types of attractors that have been studied in this context: pullback, snapshot, and uniform, as distinct from the forward attractors that are more widely known in the autonomous context; further mathematical details are provided in Appendix~A. \\
The first finding, as suspected, was that the time-invariant, forward attractor for a perpetual summer, at $F=6,$ or a perpetual winter, at $F=8,$ is not the same as a snapshot of a nonautonomous forward attractor at a time that matches the same $F$-value from a model version with seasonal forcing. \\
These comparisons were made using the heat maps of ensembles started with $5 \cdot 10^4$ randomly selected initial conditions in a cube $D \subset \mathbb{R}^3$ defined as $D = \{(X,Y,Z) \vert -3  \le X \le 3, -3 \le Y \le 3, -3 \le Z \le 3 \}.$ In the summer case, the heat map of the snapshot in Fig.~\ref{fig:PBA_F6}(b) is just slightly noisier than that of the perpetual summer in Fig.~\ref{fig:PBA_F6}(a). The comparison for the winter case shows a much more strongly reduced footprint of the heat map for the snapshot, in Fig.~\ref{fig:PBA_F8}(b), and a substantial distortion with respect to  Fig.~\ref{fig:PBA_F8}(b). \\
The most striking results were obtained when imposing a linear trend in the forcing (Figs.~\ref{fig:january_negative}--\ref{fig:statistics}) or such a trend combined with seasonal forcing (Fig.~\ref{fig:PBA_F199}). In this situation, the comparisons were made between heat maps based on an entire month of January or July, in accordance with the two-timing introduced into nonautonomous dynamics studies by \cite{Flandoli.ea.2022}: a month can be taken as the time unit of the slow time $\tau$, while a time unit of the fast time $t$ is just several days, as gotten from a nondimensional analysis of the model. \\
What is expected for the mid-latitude atmospheric circulation when changing the pole-to-equator temperature gradient is a quasi-geostrophic response with jet intensity $X$ roughly proportional to or, at least, positively correlated with the forcing intensity $F$ \citep{jetstream}. Our results are much more complex. \\
We chose to impose fairly strong trends of $\alpha = \pm 2/100$~year$^{-1}$, i.e., over a century, summer can become present-day winter, and vice versa, so as not to require integration times that are too long. Still, not much that is observable with certainty happens in 5 years, but quite noticeable changes in the heat maps do occur within 50 years, and the changes over 100 years are most impressive in looking at Figs.~\ref{fig:january_negative} and ~\ref{fig:july_negative}. \\
The change in the mean intensity of the westerlies in Fig.~\ref{fig:statistics} is not consistent from one case to another, and it shows a sharp drop for January and a negative trend (panel (e)), followed by strong oscillations and a return to the initial values at the end of the century. In parallel, the mean wave energy $E_{YZ} = Y^2 + Z^2$ does follow roughly the trend of the forcing, but the most striking changes are in its kurtosis, which falls with a strong and uniform slope in panel (b), even faster for the first 50 years in panel (f) and then flattens out, and finally drops and then rises again in panel (h). The surprising changes in kurtosis are fairly remarkable, since they imply quite significant changes in the distribution of extreme events.\\
Clearly, using the concepts and methods of nonautonomous dynamics to study the combined effect of the seasonal forcing and the climate change trend have much more to teach us. In the interests of concision, we have limited ourselves to an illustration of this combined effect in Fig.~\ref{fig:PBA_F199}. In its panel (a), the forward attractor for a perpetual summer with $F = 1.99$ for the month of July in year 150 of a negative trend shows pronounced bistability between a fixed point that attracts $98 \%$ of the orbits and a limit cycle. It the figure's panel (b), the combination of the trend and seasonal forcing only leaves the limit cycle, which attracts all the orbits. The addition of the seasonal cycle thus leads to a complete shift from a prevalent steady state to a slightly diffuse set of nearly periodic orbits.\\
Given the simplicity of the model, these results have to be taken with considerable caution and tested with more detailed climate models in the hierarchy \citep{Ghil.2001, Held.2005}. For instance, the L84 model's waves are not Rossby waves and thus the model cannot tell us much about how the occurrence and duration of blocking events might be affected by anthropogenic forcing trends, a topic of considerable current interest \citep{ghil_S2S,Ghil.Lucar.2020, jetstream}. 
Still, the present results alert us to the fact that the methods of nonautonomous dynamics emphasized herein lead to results that are quite distinct from those of the autonomous dynamics applied so far, mostly if not exclusively, to the problems of climate change. \\
Finally, from the mathematical viewpoint, we found in Appendix B that the combination of chaotic intrinsic variability with fairly general but still bounded forcing can lead to unique, albeit rather complex global attracting sets. Within such a set, local attractors can coexist or disappear. Such numerical results do certainly need, as well as encourage, more rigorous examination.

\section*{Data Availability Statement}
All the programs needed to generate the figures in the paper are on a GitHub repository at
\url{https://github.com/BernardoMaraldi/S2S-atmospheric-variability-with-Lorenz-84}.

\section*{Acknowledgements} It is a pleasure to acknowledge Eviatar Bach, Gisela D. Charó and Mickaël D. Chekroun  for helpful exchanges on the computations for nonautonomous dynamics, and Tamás Bodái, Celso Grebogi and Tamás Tél for correspondence on pullback and snapshot attractors.
B.M. would like to thank the colleagues from IMAU for insightful discussions and computational support and the staff of the Laboratoire de Météorologie Dynamique at the École Normale Supérieure for their hospitality; his work was supported by an Erasmus Fellowship. This paper is ClimTip contribution \#[XYZ]; the Quantifying climate tipping points and their impacts (ClimTip) project has received funding from the European Union’s Horizon research and innovation programme under grant agreement No. 101137601. M.G. also received support from the French Agence Nationale pour la Recherche (ANR) project TeMPlex under grant award ANR-23-CE56-1214 0002.

\section*{Appendix A. Pullback, forward and uniform attractors}
\label{app:appendix}
\begin{figure}    \includegraphics[width=0.99\columnwidth]{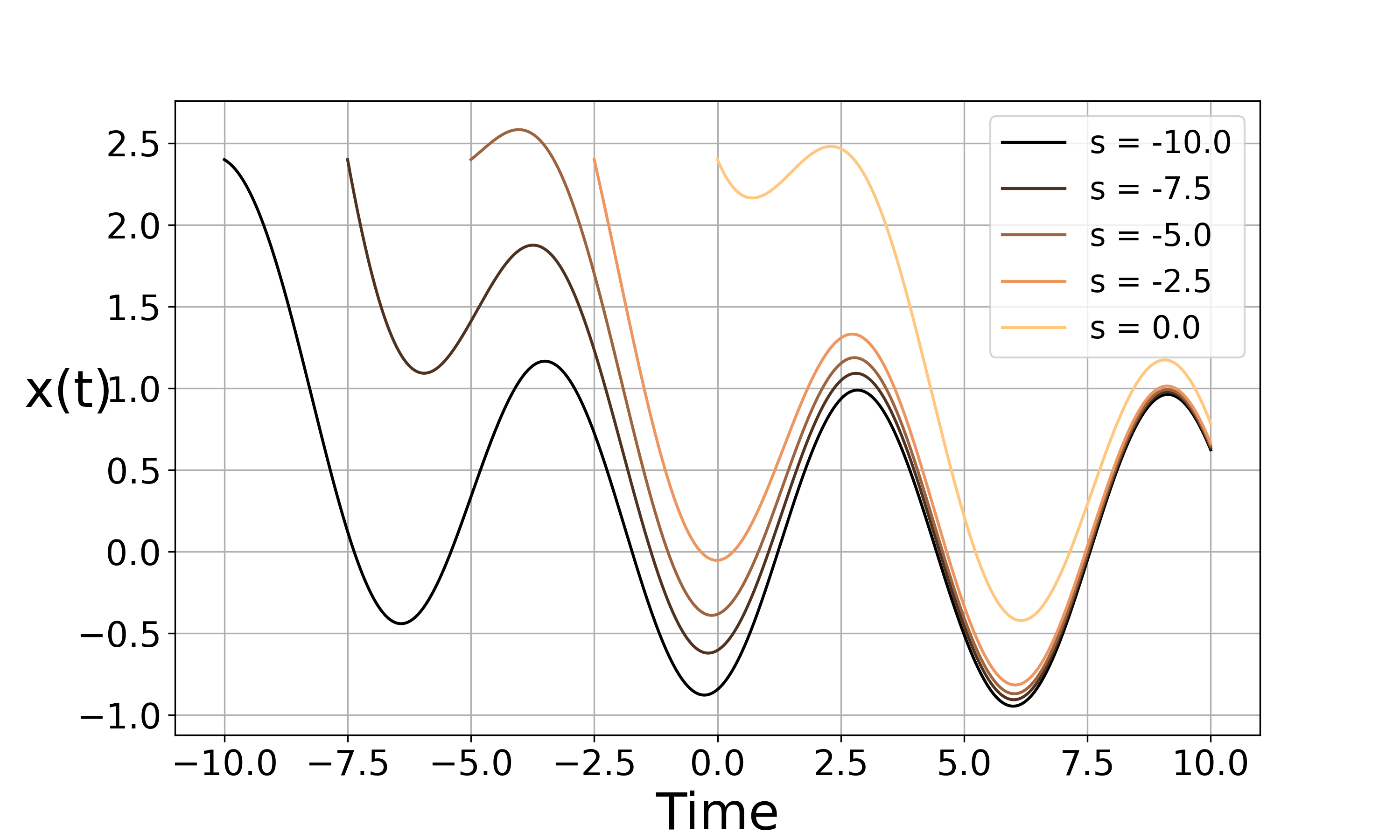}
    \caption{Solutions of Eq.~\eqref{pb}, each starting at a different initial time $s$. The convergence to the periodic PBA improves as $|s|$ increases, i.e., as one pulls back from the observation time $t$. Parameter values here are $\alpha = 0.3, b=1$ and initial value $x_0=1$. }
    \label{fig:PBA_concept}   
\end{figure}

This appendix provides a more detailed mathematical setting for the concept of pullback attractors (PBAs) in the context of nonautonomous dynamical systems, and for clarifying the relation between PBAs, on the one hand, and forward and uniform attractors, on the other. The notation follows \cite{ghil_PBA} and \cite{charo}. \\
The system under consideration is:
\begin{equation}
    \dot{\mathbf{x}}=\mathbf{f}(t, \mathbf{x}), \quad \x \in X,  \tag{A1}
    \label{eq:NDS}
\end{equation}
where $\mathbf{x}$ defines the state of the system in the phase space $X$, the dot indicates differentiation with respect to time, and $\mathbf{f}$ determines the evolution of $\x$ over time. We look for the general solution $\varphi(s,t)\mathbf{x}$ of the initial value problem given by \eqref{eq:NDS} and the initial condition $\mathbf{x}(s) = \mathbf{x}_0$. \\
For deterministic autonomous dynamical systems, i.e., when $\partial \f/\partial t \equiv \0,$ only the interval $t' = t-s$ matters for the system's evolution, since it determines completely its state at time $t$, given uniqueness of solutions for smooth $\f$ and prescribed $\x_0$. Thus, in the autonomous case, the operator $\varphi(s,t)$ provides a one-parameter description of the system's evolution,
since $s$ is arbitrary but fixed.
When the system's dynamics, though, depends explicitly on time, as in \eqref{eq:NDS}, the full solution $\varphi(s,t)$ becomes a two-parameter operator, depending on both the initial time $s$ and the observation time $t$. \\
A new type of attraction can then be defined, and the result is a new object, the PBA, which satisfies the following \\
{\bf Definition.} The indexed family of objects $\mathcal{A}=\{A(t)\}_{t\in \mathbb{R}}$, where each snapshot $A(t) \in \mathcal{A}$ is a compact subset of the phase space $X$, is a {\em pullback attractor (PBA)} if, for all $t$:
\begin{enumerate}[(i), nosep]
    \item $A(t)$ is invariant with respect to the dynamics: $\varphi (t,s) A(s)=A(t)$ for every $s\leq t$; and
    \item $\lim\limits_{s \to -\infty} \text{dist}(\varphi(s, t)B, A(t)) = 0$ for every bounded subset $B \subset X$ and $\text{dist}$ is the Hausdorff semi-distance between sets.
\end{enumerate} 
Further details can be found in \cite{Car.Han.2016} and \cite{Kloeden.Yang.2020}. \\
The random version of a PBA is called a {\em random attractor}. Its definition and study preceded actually the deterministic version above \citep{Crauel.Flan.1994} and were used in the climate literature by \cite{ghil_PBA} and \cite{chekroun}. Since the concepts and methods involved are more complicated than those required in the deterministic case treated herein, we do not present them in this appendix. \\
In the physical literature, a {\em snapshot attractor} is simply defined as ``the pattern formed by a cloud of orbits at a fixed time" when the forcing is time dependent \citep{Namenson.ea.1996} and it does not require an explicit pullback $s \to - \infty$. This looser concept has been effectively used in the climate literature as well \citep{drotos, Tel.ea.2020}. The snapshot attractor concept corresponds in the mathematical literature to the generalization of a forward attractor to time-dependent forcing \citep{Car.Han.2016, Kloeden.Yang.2020}. We have chosen to retain the name of snapshot attractor here, rather than the longer mathematical formulation of "nonautonomous forward attractor." \\
Forward attraction, though, becomes more difficult to apply in a nonautonomous case as simple as 
\begin{equation} \label{eq:monotone}
\dot x = - \alpha x + \sigma t, \quad x(0) = x_0, \tag{A2}
\end{equation}
where both $ \alpha > 0$ and  $ \sigma > 0$. In this case, there are no bounded solutions for $t >0$ and no limit sets in the usual forward sense. Hence there seems to be no good definition in the snapshot sense of what a future climate might look like if anthropogenic greenhouse gas emissions keep rising. The pullback definition, though, yields the simple straight-line PBA $A(t)$ given by
\begin{equation}
	{A(t)} = - \frac{\sigma}{\alpha} \left( t - \frac{1}{\alpha} \right); \tag{A3}
\end{equation}
see \cite{Ghil.RDS.2021} for details and a graphic illustration. \\
A forced and dissipative system for which it is immediate to show that forward and pullback attraction both work and how they differ is the following:
\begin{equation}
    \dot x = - \alpha x + b \sin t ,
    \tag{A4}
    \label{pb}
\end{equation}
with the initial condition \( x(s) = x_0 \), where \( t \geq s \). The limit for $t \to + \infty$ and $s$ fixed is discussed in \citet[Sec.~3.2.1]{Car.Han.2016} and illustrated there in Fig.~3.2. \\
On the other hand, the limit for $s \to - \infty$ is well defined in the pullback sense above, and it is illustrated in Fig.~\ref{fig:PBA_concept} here and in  \citet[Fig.~3.1]{Car.Han.2016}.  
Namely the difference between any two solutions will decrease as $s \to - \infty$ and the particular solution to which the common behavior of all solutions tends as $s \to - \infty$ is given by 
\begin{equation}
	{A(t)} = - \frac{b(\alpha \sin t - \cos t)}{a^2 + 1} , \tag{A5}
	\label{PBA:per}
\end{equation}
which is the corresponding PBA \citep{Car.Han.2016}. \cite{keno} provide a similar PBA example for periodic forcing that is relevant to the orbital forcing of the Quaternary glaciations. \\
Cases in which forward attraction can be defined — although the forcing is no longer purely periodic but still bounded — exist and are described with mathematical rigor by \citet[Ch.~10--12]{Kloeden.Yang.2020}. The condition of boundedness of the forcing, though, seems to be necessary for snapshot attractors to exist. Furthermore, even when they do exist, they may not be unique. \\
When the limit (ii) in the Definition of a PBA above is uniform in $t$ and a similar property for the equivalent definition of a forward attractor holds with respect to $s$, one speaks of a {\em uniform attractor}. For such an attractor, the two concepts, of forward and pullback attraction, hold and are equivalent. This is the case for more general systems under periodic forcing than that of Eq.~\eqref{pb} \citep{Car.Han.2016, Kloeden.Yang.2020}. It is also true, in particular, for the seasonal forcing of the L84 model \citep{Anguiano.Car.2014}.\\ 
\cite{Haraux.1991} and \cite{Vishik.1992} developed a theory of uniform attractors, which does apply for the seasonal forcing of the L84 model \citep{Anguiano.Car.2014, drotos} but not to a monotonically increasing forcing that imitates anthropogenically unbounded changes in greenhouse gas emissions.\\ 
Clearly, the pullback limit $s \to - \infty$ is not achievable in a finite-time numerical integration. When computing an approximate PBA in practice, it depends on the parameters of the problem --- in particular the dissipativity, which is given by $\alpha$ in Eqs.~\eqref{eq:monotone} and \eqref{PBA:per} above --- how far one has to "pull back." In many practical problems, like in Fig.~\ref{fig:PBA_concept}, it suffices to pull back a small multiple of the characteristic times of the model \citep{chekroun, Pierini.Ghil, charo}. 

\section*{Appendix B. Dissipativity and globally attracting set}
Consider the system~\eqref{eq:L84}, which we rewrite for convenience here as
\begin{subequations} 
	\begin{align}
		& \dot X = -Y^2 -Z^2 - aX + aF, \tag{B1.a} \\
		& \dot Y = XY - bXZ - Y + G, \tag{B1.b} \\
		&  \dot Z = bXY + XZ - Z,  \tag{B1.c}
	\end{align}
	\label{eq:Lor84}
\end{subequations}
where $\dot X \equiv {d X}/{d t}$ and so on. Note that this system has the general form 
\begin{equation} \label{eq:FD}
	\dot X_i = A_{ijk} X_j X_k - B_{ij} X_j + C_i, \quad i = 1,2, \ldots N, \tag{B2}
\end{equation}
where we use the summation convention for repeated indices. Here $\X = (X_1, X_2, \ldots, X_N)$ is an $N$-vector, while $\A = ( A_{ijk}), \B =  (B_{ij})$ and $\C = (C_i)$ are constant tensors of appropriate dimensions, with $1 \le i,j,k \le N.$ \\
In the present case, like for the \cite{Lorenz.1963} convection model and for many other models from fluid dynamics and elsewhere, the coefficients $( A_{ijk}), (B_{ij})$ satisfy the following two important conditions,
\begin{subequations} \label{eq:C+D}
	\begin{align}
		& A_{ijk} X_i X_j X_k = 0, \tag{C} \\ 
		& B_{ij} X_i X_j > 0, \tag{D} 
	\end{align}
\end{subequations}
which hold for arbitrary values of the $X_i$'s, with $\X \neq 0$ in (D).\\
Property (C) states that, in the absence of the other terms, $\B = 0, \; \C = \0,$ system (B1) would be conservative. Indeed, defining the energy as $E = (X^2+ Y^2 + Z^2)/2,$ and multiplying each of the equations in \eqref{eq:L84} by $X, Y$ and $Z$, respectively, one has $\dot E = 0,$ i.e., $E =$~const. \\
Property (D) is equivalent to the matrix $\mathcal B$ being positive definite. Physically, it means that the system contains a dissipative mechanism. Indeed, for the full system, 
\begin{equation} \label{eq:diss}
	\dot E = - B_{ij} X_i X_j + C_i X_i. \tag{B3}
\end{equation}
This shows that $\B$ causes the energy of the system to decrease, if (D) holds. \\
The rate of change $\dot E$ of $E(\X)$ along a system trajectory $\X = \X(t)$ is given by \eqref{eq:diss}. On any energy surface in phase space, $E = $~const., this rate attains its algebraically largest value for $\X$ pointing in the direction of the vector $\C$, which achieves 
$\max \{C_i X_i\} = cx$, where $c^2 = C_i C_i$ and $x^2 = X_i X_i = 2E$. Thus
\begin{equation}
\dot E \le - \beta_1 x^2 + cx,   \tag{B4}
\end{equation}
where $\beta_1 = \min_{\{x=1\}} {B_{ij} X_i X_j > 0}$ is the lowest eigenvalue of the matrix $\B$. Clearly $ \dot E = x \dot x$ will be negative for any $x > c/\beta_1.$ 

It follows that trajectories of a forced-dissipative system \eqref{eq:FD} that satisfies conditions (C) and (D) will all eventually enter a ball $\mathscr B = \{X: x \le c/\beta_1\},$ never to leave it again. In the absence of forcing, $\C = \0$, this ball is reduced to a point, the origin $\X = \0.$ It is the competition between the forcing $\C$ and the dissipation $\B$ that renders the behavior of these systems interesting. Part of the interest, both physical and mathe­matical, derives from the fact that this competition has to be played out within the bounded ball $\mathscr B.$ \\
In particular, for system (B1), 
$\B$ is diagonal with elements $\{a, 1, 1\}$ on the diagonal and $\beta_1 = \min\{a,1\},$ while $C = (aF, G, 0)$. Hence the globally attracting set for the forward orbits of the autonomous system is the ball $\mathscr B$ of radius $ r = (a^2 F^2 + G^2)^{1/2}/\min\{a,1\}.$ 
A typical radius then, for $F = 6, a = 1,$ and $G \equiv 1$ is $r = \sqrt 37 \simeq 6.1.$ In the autonomous case, the two limit cycles of Fig.~\ref{fig:basins}(a) fit easily into this much larger ball. \\
The more interesting case is that of a time-dependent $F$, especially one for which the autonomous attractor is not just a fixed point, and for which $F$ is not purely periodic or a mild generalization thereof, like the cases treated in \citet[Figs.~1--6]{Anguiano.Car.2014}. Examining Fig.~\ref{fig:statistics}, we see that the means (black line) of both the jet intensity $X$ and of the wave energy $E = Y^2 + Z^2$ in the figure's two columns hover close to a value of 1. Strikingly, this is even so in panel (e), in which $X(t)$ undergoes sharp fluctuations.

\section*{References}
\nocite{*}
 \bibliography{Maraldi.ea-MG2}

\end{document}